\documentclass[mssymb,epsfig]{article}
\pdfoutput=1
\usepackage{graphicx}
\usepackage{bm}
\usepackage{fullpage}
\usepackage{amsmath}
\usepackage[pdftex]{color}

\usepackage{bm}
\usepackage{color}

\def\lsim{\mathrel{\mathop
  {\hbox{\lower0.5ex\hbox{$\sim$}\kern-0.8em\lower-0.7ex\hbox{$<$}}}}}
\def\gsim{\mathrel{\mathop
  {\hbox{\lower0.5ex\hbox{$\sim$}\kern-0.8em\lower-0.7ex\hbox{$>$}}}}}

\begin{document}

\newcommand{\1}{$\spadesuit$}
\newcommand{\half}{{1\over2}}
\newcommand{\nad}{n_{\rm ad}}
\newcommand{\niso}{n_{\rm iso}}
\newcommand{\ncor}{n_{\rm cor}}
\newcommand{\fiso}{f_{\rm iso}}
\newcommand{\ii}{\'{\'i}}
\newcommand{\bk}{{\bf k}}
\newcommand{\Ocdm}{\Omega_{\rm cdm}}
\newcommand{\ocdm}{\omega_{\rm cdm}}
\newcommand{\OM}{\Omega_{\rm M}}
\newcommand{\OB}{\Omega_{\rm B}}
\newcommand{\oB}{\omega_{\rm B}}
\newcommand{\OX}{\Omega_{\rm X}}
\newcommand{\cltt}{C_l^{\rm TT}}
\newcommand{\clte}{C_l^{\rm TE}}
\newcommand{\calR}{{\cal R}}
\newcommand{\calS}{{\cal S}}
\newcommand{\Rrad}{{\cal R}_{\rm rad}}
\newcommand{\Srad}{{\cal S}_{\rm rad}}
\newcommand{\calPR}{{\cal P}_{\cal R}}
\newcommand{\calPS}{{\cal P}_{\cal S}}

\input epsf

\title{An analytical approach to Bayesian evidence computation}
\author{Juan Garc{\'\i}a-Bellido \\
Departamento de F\'\i sica Te\'orica \ C-XI, Universidad
Aut\'onoma de Madrid, \\ Cantoblanco, 28049 Madrid, Spain}
\date{April 14th, 2005}

\maketitle

\begin{abstract}
The Bayesian evidence is a key tool in model selection, allowing a comparison of 
models with different numbers of parameters. Its use in analysis of cosmological 
models has been limited by difficulties in calculating it, with current 
numerical algorithms requiring supercomputers. In this paper we give exact 
formulae for the
Bayesian evidence in the case of Gaussian likelihoods with arbitrary
correlations and top-hat priors, and approximate formulae for the case of
likelihood distributions with leading non-Gaussianities (skewness and
kurtosis). We apply these formulae to cosmological models with and without 
isocurvature components, and compare with results we previously obtained using 
numerical thermodynamic integration. We find that the results are of lower 
precision than the thermodynamic integration, while still being good enough to 
be useful.
\end{abstract}

\section{Introduction}

Model selection refers to the statistical problem of deciding which
model description of observational data is the best \cite{jeff,mackay}. It 
differs from
parameter estimation, where the choice of a single model (i.e.~choice of 
parameters
to be varied) has already been made and the aim is to find their best-fitting 
values and ranges. While there have been widespread
applications of parameter estimation techniques, usually likelihood
fitting, to cosmological data, there has so far been quite limited
application of model selection statistics 
\cite{evidence,Liddle:2004nh,Beltran2005}. This is unfortunate, as
model selection techniques are necessary to robustly distinguish
between models with different numbers of parameters, and many of the
most interesting issues in cosmology concern the desirability or
otherwise of incorporating additional parameters to describe new
physical effects.

Within the context of Bayesian inference, model selection should be
carried out using the {\em Bayesian evidence} \cite{jeff,mackay}, which measures 
the probability of the model in light of the observational data (i.e.~the
average likelihood over the prior distribution). The Bayesian evidence
associates a single number with each model, and the models can then be
ranked in order of the evidence, with the ratios of those values
interpretted as the relative probability of the models. This process
sets up a desirable tension between model simplicity and ability to
fit the data.

Use of the Bayesian evidence has so far been limited by difficulties
in calculating it. The standard technique is thermodynamic
integration \cite{thermo,Hobson2002}, which varies the temperature in a Monte 
Carlo Markov
Chain (MCMC) approach in order that the distribution is sampled in a
way covering both posterior and prior distributions. However, in
recent work \cite{Beltran2005} we showed that in order to obtain
sufficiently-accurate results in a cosmological context, around $10^7$
likelihood evaluations are required per model. Such analyses are
CPU-limited by the time needed to generate the predicted spectra to
compare with the data, and this requirement pushes the problem into
the supercomputer class (for comparison, parameter estimation runs
typically employ $10^5$ to $10^6$ likelihood evaluations). 

In this paper, we propose and exploit a new analytic method to compute
the evidence based on an expansion of the likelihood distribution
function. The method pre-supposes that the covariance of the posterior
distribution has been obtained, for instance via an MCMC parameter
estimation run, and in its present form requires that the prior
distributions of the parameters are uniform top-hat
priors.\footnote{An extension to gaussian priors should be feasible,
but not one to arbitrary priors.}  While the method will not be
applicable for general likelihood distributions, we include the
leading non-gaussianities (skewness and kurtosis) in approximating the
likelihood shape, with the expectation of obtaining good results
whenever the likelihood distribution is sufficiently
simple. Cosmological examples commonly exhibit likelihood
distributions with only a single significant peak.

We apply the method both to toy model examples and to genuine
cosmological situations. In particular, we calculate the evidences for
adiabatic and isocurvature models, which we previously computed using
thermodynamic integration in Ref.~\cite{Beltran2005}. We find that the 
discrepancies between the methods are typically no worse than 1 in ln(Evidence), 
meaning that the analytic method is somewhat less accurate than would be ideal, 
but is accurate enough to give a useful indication of model preference.

\section{The Bayesian evidence}

The posterior probability distribution ${\cal P}({\bf \theta},{\cal
M}|{\bf D})$ for the parameters ${\bf \theta}$ of the model ${\cal
M}$, given the data ${\bf D}$, is related to the likelihood function
${\cal L}({\bf D}|{\bf \theta},{\cal M})$ within a given set of prior
distribution functions $\pi({\bf \theta},{\cal M})$ for the parameters
of the model, by Bayes' theorem:
\begin{equation}\label{Posterior}
{\cal P}({\bf \theta},{\cal M}|{\bf D}) = {{\cal L}({\bf D}|{\bf
\theta},{\cal M})\,\pi({\bf \theta},{\cal M})\over E({\bf D}|{\cal M})}\,,
\end{equation}
where $E$ is the Bayesian evidence, i.e. the average likelihood over
the priors, 
\begin{equation}\label{evidence}
E({\bf D}|{\cal M}) = \int d{\bf \theta}\ {\cal L}({\bf D}|{\bf \theta},
{\cal M})\,\pi({\bf \theta},{\cal M})\,,
\end{equation}
where ${\bf \theta}$ is a vector with $n$-components characterising
the $n$ independent parameters. The prior distribution function $\pi$
contains all the information about the parameters before observing the
data, i.e. our theoretical prejudices, our physical understanding of
the model, and input from previous experiments. 

In the case of a large number of parameters ($n\gg 1$), the evidence
integral cannot be performed straightforwardly and must be obtained
either numerically or via an analytic approximation. Amongst numerical
methods the most popular is thermodynamic integration
\cite{thermo,Hobson2002} but this can be computationally extremely
intensive \cite{Beltran2005}. The simplest analytical approximation is
the Laplace approximation, valid when the distribution can be
approximated by a multivariate Gaussian. This may hold when the
quantity and quality of the data is optimal, but is likely to be valid
only in limited cosmological circumstances.

The Bayesian evidence is of interest because it allows a comparison of
models amongst an exclusive and exhaustive set $\lbrace {\cal{M}}_i 
\rbrace_{i=1...N}$. We can compute
the posterior probability for each hypothesis given the data ${\bf D}$ using 
Bayes theorem:
\begin{equation}\label{Bayes}
{\cal P}({\cal M}_i|{\bf D}) \propto E({\bf D}|{\cal M}_i)\,
\pi({\cal M}_i)\,,
\end{equation}
where $E({\bf D}|{\cal M}_i)$ is the evidence of the data under the
model ${\cal M}_i$, and $\pi({\cal M}_i)$ is the prior probability of
the $i$th model {\em before} we see the data. The ratio of the
evidences for the two competing models is called the {\em Bayes
factor} \cite{KR95}
\begin{equation}\label{BayesFactor}
B_{ij} =  {E({\bf D}|{\cal M}_i)\over E({\bf D}|{\cal M}_j)}\,,
\end{equation}
and this is also equal to the ratio of the posterior model probabilities if we
assume that we do not favour any model \textit{a priori},
so that $\pi({\cal M}_1)=\pi({\cal M}_2)=...=\pi({\cal M}_N)=1/N$. 

The Bayes factor Eq.~(\ref{BayesFactor}) provides a
mathematical representation of Occam's razor, because more complex models tend 
to be less predictive, lowering their average likelihood in comparison to 
simpler, more predictive models. More complex models can only be favoured if 
they are able to provide a significantly improved fit to the data. In simple 
cases where models give vastly different maximum likelihoods there is no need to 
employ model selection techniques, but they are essential for properly 
discussing cases where the improvement of fit is marginal. This latter situation 
is more or less inevitable whenever the possibility of requiring an additional 
parameter arises from new data, unless the new data is of vastly greater power 
than that preceding it; cosmological examples include the inclusion of spectral 
tilt, dark energy density variation, or the case explored later in this paper of 
trace isocurvature perturbations.

In this paper we will obtain an analytical formula which approximates
the Bayesian evidence by considering the higher-order cumulants of the 
distribution in a systematic way. The advantage is that with
these analytical formulae one can compute the evidence for a given
model with an arbitrary number of parameters, given the hierarchy of
cumulants of the distribution, assumed previously computed for the 
likelihood distribution function within the parameter estimation
programme.

The evidence needs to be calculated to sufficient precision for robust 
conclusions to be drawn. The standard interpretational scale, due to Jeffreys 
\cite{jeff} and summarized in Ref.~\cite{Beltran2005}, strengthens its verdict 
roughly each time the difference in ln(Evidence) increases by one. The evidence 
therefore needs to be computed more accurately than this, with an uncertainty of 
0.1 in ln(Evidence) easily sufficient, and a factor two worse than that 
acceptable. This accuracy requirement ensures that the relative model 
probabilities are little changed by the uncertainty.

The first thing we need is to characterize the distribution function
for the model with $n$ parameters. Let $f({\bf x})$ be this function,
and let us assume that it is properly normalized,
\begin{equation}\label{norm}
\int_{-\infty}^{\infty} d^n{\bf x}\, f({\bf x}) = 1\,.
\end{equation}
Then, the $p$-point correlation function is given by
\begin{equation}\label{pcorr}
\langle x_{i_1}\dots x_{i_p}\rangle = 
\int_{-\infty}^{\infty} d^n{\bf x}\ x_{i_1}\dots x_{i_p}\,f({\bf x})\,.
\end{equation}
{}From this distribution function one can always construct the generating
functional, $\phi({\bf u})$, as the Fourier transform
\begin{equation}\label{phiFT}
\phi({\bf u}) =
\int_{-\infty}^{\infty} d^n{\bf x}\, e^{i\,{\bf u}\cdot{\bf x}}\,
f({\bf x})\,.
\end{equation}
This function can be expanded as
\begin{equation}\label{phiS}
\phi({\bf u}) = \exp\left[\sum_{p=1}^\infty {i^p\over p!}\,
A_{i_1\dots i_p}\,u^{i_1}\dots u^{i_p}\right]\,,
\end{equation}
where $A_{i_1\dots i_p}$ are totally symmetric rank-$p$ tensors.
For instance, if we restrict ourselves to order 4, we can write
\begin{equation}\label{phi}
\phi({\bf u}) = \exp\left[i\,\mu_i u_i - {1\over2!}\, C_{ij}
\,u_i u_j - \,{i\over3!}\, B_{ijk}\,u_i u_j u_k +
\,{1\over4!}\, D_{ijkl}\,u_i u_j u_k u_l + \dots +
\,{i^n\over n!}\, A_{i_1\dots i_n}\,u_{i_1}\dots u_{i_n} \right]\,,
\end{equation}
where $\mu_i$ is the mean value of variable $x_i$; $ C_{ij}$ is the
covariance matrix; $B_{ijk}$ is the trilinear matrix associated with
the third cumulant or skewness; $D_{ijkl}$ is the rank-4 tensor
associated with the fourth cumulant or kurtosis, and $A_{i_1\dots i_n}$ 
is the rank-$n$ tensor associated with the $n$-th cumulant. Their expressions
in terms of $n$-point correlation functions can be obtained from
Eq.~(\ref{phiFT}), by realising that
\begin{equation}\label{ncorr}
\langle x_{i_1}\dots x_{i_n}\rangle = (-i)^n\left.
{\partial^n\phi({\bf u})\over\partial u_{i_1}\dots\partial u_{i_n}}
\right|_{{\bf u}=0}\,.
\end{equation}
For instance, the first-order term gives
\begin{equation}
\langle x_i\rangle = (-i)\left.
{\partial\phi({\bf u})\over\partial u_i}\right|_{{\bf u}=0}=\mu_i\,.
\end{equation}
The second-order correlation function gives
\begin{equation}
\langle x_i x_j\rangle = (-i)^2\left.{\partial^2\phi({\bf u})\over
\partial u_i\partial u_j}\right|_{{\bf u}=0}= C_{ij}+\mu_i\mu_j\,,
\end{equation}
such that the covariance matrix is obtained, as usual, from
$$ C_{ij} = \langle x_i x_j\rangle - \langle x_i\rangle\langle x_j
\rangle\,.$$
The third-order correlation function gives
\begin{equation}
\langle x_i x_j x_k\rangle = (-i)^3\left.{\partial^3\phi({\bf u})\over
\partial u_i\partial u_j\partial u_k}\right|_{{\bf u}=0}=  B_{ijk}+
\mu_i  C_{jk} + \mu_j  C_{ki} + \mu_k  C_{ij} + \mu_i\mu_j\mu_k\,,
\end{equation}
such that the skewness matrix is obtained from
\begin{equation}
 B_{ijk} = \langle x_i x_j x_k\rangle - \langle x_i\rangle
\langle x_j x_k\rangle - \langle x_j\rangle
\langle x_k x_i\rangle - \langle x_k\rangle
\langle x_i x_j\rangle + 2\langle x_i\rangle\langle x_j\rangle
\langle x_k\rangle \,.
\end{equation}
The fourth-order correlation function gives
\begin{eqnarray}
\langle x_i x_j x_k x_l\rangle = (-i)^4\left.{\partial^4\phi({\bf
u})\over \partial u_i\partial u_j\partial u_k\partial u_l}
\right|_{{\bf u}=0} &\!=\!&  D_{ijkl} +  C_{ij} C_{kl} + 
 C_{ik} C_{jl} +  C_{il} C_{jk} \\ \nonumber
&\!+\!& B_{ijk}\mu_l +  B_{ijl}\mu_k +  B_{jkl}\mu_i + 
 B_{ikl}\mu_j \\[2mm] \nonumber
&\!+\!& C_{ij}\mu_k\mu_l +  C_{ik}\mu_j\mu_l + 
 C_{il}\mu_j\mu_k \\[2mm] \nonumber
&\!+\!& C_{jk}\mu_i\mu_l +  C_{jl}\mu_i\mu_k + 
 C_{kl}\mu_i\mu_j \\[2mm] \nonumber
&\!+\!&\mu_i\mu_j\mu_k\mu_l\,, 
\end{eqnarray}
such that the kurtosis matrix is obtained from
\begin{eqnarray}
 D_{ijkl} &\!=\!& \langle x_i x_j x_k x_l\rangle - 
\langle x_i x_j\rangle\langle x_k x_l\rangle -
\langle x_i x_k\rangle\langle x_j x_l\rangle -
\langle x_i x_l\rangle\langle x_j x_k\rangle \\[2mm]\nonumber
&\!-\!&\langle x_i x_j x_k\rangle\langle x_l\rangle -
\langle x_i x_j x_l\rangle\langle x_k\rangle -
\langle x_i x_k x_l\rangle\langle x_j\rangle -
\langle x_j x_k x_l\rangle\langle x_i\rangle \\[2mm]\nonumber
&\!+\!&2\,\langle x_i x_j\rangle\langle x_k\rangle\langle x_l\rangle +
2\,\langle x_i x_k\rangle\langle x_j\rangle\langle x_l\rangle +
2\,\langle x_i x_l\rangle\langle x_j\rangle\langle x_k\rangle +
2\,\langle x_j x_k\rangle\langle x_i\rangle\langle x_l\rangle 
\\[2mm]\nonumber
&\!+\!&2\,\langle x_j x_l\rangle\langle x_i\rangle\langle x_k\rangle +
2\,\langle x_k x_l\rangle\langle x_i\rangle\langle x_j\rangle -
6\,\langle x_i\rangle\langle x_j\rangle\langle x_k\rangle
\langle x_l\rangle \,,
\end{eqnarray}
and so on, for the higher order cumulants.

\section{The Gaussian approximation}

Let us first evaluate the evidence for a multivariate Gaussian
distribution, that is, one in which all the cumulants are zero except
the covariance matrix $ C_{ij}$ and the means $\mu_i$.  In this case,
the generating functional and the distribution are given by
\begin{eqnarray}
&&\phi({\bf u}) = \exp\Big[- i \mu_i u_i-\half\, C_{ij}
\,u_i u_j\Big]\,,\\[2mm]
&&f({\bf x}) = {1\over(2\pi)^n} \int_{-\infty}^\infty d^n {\bf u}\
e^{-i\,{\bf u}\cdot{\bf x}}\ \phi({\bf u}) \\[2mm]
&&\hspace{1cm} ={1\over(2\pi)^{n/2}\sqrt{\det C}} \exp\Big[
-\half C_{ij}^{-1}(x_i-\mu_i)(x_j-\mu_j)\Big]\,,
\end{eqnarray}
which satisfies
\begin{equation}
\langle x_i\rangle = \mu_i\,, \hspace{1cm}
\langle x_i x_j\rangle =  C_{ij} + \mu_i\mu_j\,, \hspace{1cm}
\langle x_i x_j x_k\rangle = \mu_{(i} C_{jk)} + \mu_i\mu_j\mu_k\,, 
\hspace{5mm}\dots
\end{equation}
where the subindices in parenthesis, $(ijk)$, indicate a cyclic sum.
Notice that all the n-point correlation functions can be written in
terms of the first two moments of the distribution, and all the higher-order 
cumulants vanish.

\subsection{Centred priors}

For initial calculations, we assume a top-hat prior and make the unrealistic 
assumption, to be lifted later, that it is centered at the
mean value:
\begin{equation}\label{prior}
\pi(x,a) \equiv \left\{\begin{array}{ll}(2a)^{-1}&\hspace{1cm}
-a < x-\mu < a\,,\\[2mm] 0 &\hspace{1cm} {\rm otherwise}\,.\end{array}
\right.
\end{equation}
Since the Fourier transform of a top-hat function is
$$\int_{-\infty}^{\infty} dx\, e^{iux}\,\pi(x,a) = 
{\sin au\over au}\,\exp[i\mu u]\,,$$
we can write the evidence either way
\begin{eqnarray}\label{evx}
E(a_1,\dots,a_n) &\!=\!&
\int_{-\infty}^{\infty} d^n{\bf x}\,f({\bf x})\,
\prod_{i=1}^n \pi(x_i,a_i) \ = \ \prod_{i=1}^n (2a_i)^{-1} \!
\int_{-a_1}^{a_1}dx_1\dots\int_{-a_n}^{a_n} dx_n\,f(\tilde{\bf x})
 \\[3mm] \label{evu}
&\!=\!&{\displaystyle {1\over(2\pi)^n}
\int_{-\infty}^{\infty} d^n{\bf u}\,\phi({\bf u})\,
\prod_{i=1}^n {\sin a_iu_i\over a_iu_i}\,.}
\end{eqnarray}
In Eq.~(\ref{evx}) we integrate over the displaced coordinate, $\tilde x_i
\equiv x_i - \mu_i$, such that $\langle \tilde x_i\rangle = 0$ and
$\langle \tilde x_i\tilde x_j\rangle =  C_{ij}$.  From now on, we ignore the 
tildes, and assume we have moved to those coordinates.
Note that the choice of prior is not crucial. We could have chosen a
Gaussian prior, and the result would not be very different, except
that the window functions, $\sin z/z$, would then be Gaussians.  Let
us now perform the integration Eq.~(\ref{evx}) in the case of 1, 2 and
then $n$ variables.

\vspace{5mm}
{\bf 1 variable}. Suppose the covariance is just $ C=\sigma^2$.
The evidence is then
\begin{equation}\label{evidence1}
E(a) = {1\over2a\,\sigma\sqrt{2\pi}} \int_{-a}^a dx\ 
e^{-{x^2\over2\sigma^2}} = {1\over2\pi} \int_{-\infty}^{\infty} du\,
{\sin au\over au}\,e^{-\half \sigma^2 u^2} = {1\over2a}{\rm Erf}\Big[
{a\over\sigma\sqrt2}\Big]\,,
\end{equation}
where ${\rm Erf}[x]$ is the error function, which asymptotes very
quickly to one for $x\geq2$, or $a\geq3\sigma$. Therefore, the evidence
of a model with centred top-hat prior of width $2a$ is well approximated
by $(2a)^{-1}$. The wider is the theoretical prior, the smaller is the
evidence, as expected.

\vspace{5mm} {\bf 2 variables}. Suppose we have two correlated 
variables, $x_1$ and $x_2$, with covariance matrix
\begin{equation}\label{corre2}
 C = \left(\begin{array}{ll} C_{11}& C_{12}\\[2mm]
 C_{12}& C_{22}\end{array}\right) =
\left(\begin{array}{cc}\sigma_1^2&\rho\sigma_1\sigma_2\\[2mm]
\rho\sigma_1\sigma_2&\sigma_2^2\end{array}\right) \,.
\end{equation}
where the cross-correlation $\rho$ is defined by 
$$\rho = {\langle x_1x_2\rangle\over\sqrt{\langle x_1^2\rangle
\langle x_2^2\rangle}} = {\langle x_1x_2\rangle\over\sigma_1\sigma_2}\,,$$
with $\sigma_1$ and $\sigma_2$ the corresponding quadratic dispersions.
In this case, the normalized 2-dimensional distribution function is
\begin{equation}\label{f2}
f({\bf x}) = {1\over2\pi\sigma_1\sigma_2\sqrt{1-\rho^2}}\,\exp\Big[
{-1\over1-\rho^2}\Big({x_1^2\over2\sigma_1^2}-{\rho x_1x_2\over\sigma_1
\sigma_2}+{x_2^2\over2\sigma_2^2}\Big)\Big]\,,
\end{equation}
which has the property that integrating (``marginalizing'') over one of
the two variables, leaves a properly-normalized Gaussian distribution
for the remaining variable,
\begin{equation}\label{marginalization}
\int_{-\infty}^\infty dx_2\ f({\bf x}) = {1\over\sigma_1\sqrt{2\pi}}\,
e^{-{x_1^2\over2\sigma_1^2}}\,.
\end{equation}

Let us now evaluate the evidence Eq.~(\ref{evx}) by integrating first over 
the prior in $x_2$,
\begin{equation}\label{int2}
{1\over2a_2} \int_{-a_2}^{a_2} dx_2\ f({\bf x}) = 
{e^{-{x_1^2\over2\sigma_1^2}}\over\sigma_1\sqrt{2\pi}}\cdot{1\over4a_2}
\left[{\rm Erf}\Big[{a_2\sigma_1+\rho\sigma_2\,x_1\over\sigma_1\sigma_2
\sqrt{2(1-\rho^2)}}\Big] + {\rm Erf}\Big[{a_2\sigma_1-\rho\sigma_2\,x_1
\over\sigma_1\sigma_2\sqrt{2(1-\rho^2)}}\Big]\right]\,.
\end{equation}
The first term is the result we would have obtained if we had been
marginalizing over $x_2$; the second is a sum of error functions that
still depend on $x_1$, and modulates the marginalization.  We can use
the series expansion of the error function to second order,
$${1\over2}\Big({\rm Erf}[a+x] + {\rm Erf}[a-x]\Big) = 
{\rm Erf}[a] - {2a\,x^2\over\sqrt\pi}\,e^{-a^2}+{\cal O}(x^4)\,,$$
to write Eq.~(\ref{int2}) to order $x_1^2$ as
\begin{equation}\label{int2b}
{1\over2a_2} \int_{-a_2}^{a_2} dx_2\ f({\bf x}) = 
{e^{-{x_1^2\over2\sigma_1^2}}\over\sigma_1\sqrt{2\pi}}\left[
{1\over2a_2}\,{\rm Erf}\Big[{a_2\over\sigma_2\sqrt{2(1-\rho^2)}}\Big] -
{\rho^2\,x_1^2\,e^{-{a_2^2\over2\sigma_2^2(1-\rho^2)}}
\over2\sigma_1^2\sigma_2(1-\rho^2)\sqrt{2\pi(1-\rho^2)}}\right]\,.
\end{equation}
Integrating now over the $x_1$ prior, we finally obtain the evidence
\begin{eqnarray}\nonumber
E(a_1,a_2) &\!=\!& {1\over4a_1a_2} \int_{-a_1}^{a_1} dx_1
\int_{-a_2}^{a_2} dx_2\ f({\bf x}) \\[2mm] \label{int12}
&\!=\!& {1\over4a_1a_2}\,
{\rm Erf}\Big[{a_2\over\sigma_2\sqrt{2(1-\rho^2)}}\Big]
{\rm Erf}\Big[{a_1\over\sigma_1\sqrt{2}}\Big] \\[2mm] \nonumber
&\!-\!& {\rho^2\,e^{-{a_2^2\over2\sigma_2^2(1-\rho^2)}}
\over2\sigma_1\sigma_2(1-\rho^2)\sqrt{2\pi(1-\rho^2)}}\,
{{\rm Erf}\Big[{a_1\over\sigma_1\sqrt{2}}\Big]\over2a_1} +
{\rho^2\,e^{-{a_2^2\over2\sigma_2^2(1-\rho^2)}-
{a_1^2\over2\sigma_1^2}}\over4\pi\sigma_1^2\sigma_2\sqrt{1-\rho^2}}\,.
\end{eqnarray}

Note that in the limit of no cross-correlations, $\rho\to0$, the
integral factorizes and we can write an exact expression for the
evidence,
\begin{eqnarray}
E(a_1,a_2) &\!=\!&{1\over4a_1a_2}\,{1\over2\pi\sigma_1\sigma_2} 
\int_{-a_1}^{a_1} dx_1 \int_{-a_2}^{a_2} dx_2\ 
e^{-{x_1^2\over2\sigma_1^2}-{x_2^2\over2\sigma_2^2}} \\[2mm]
&\!=\!&{1\over4\pi^2} \int_{-\infty}^{\infty} du_1
\int_{-\infty}^{\infty} du_2\,
{\sin a_1u_1\over a_1u_1}\,{\sin a_2u_2\over a_2u_2}\,
e^{-\half \sigma_1^2 u_1^2-\half \sigma_2^2 u_2^2} \\[2mm]
&\!=\!&{1\over4a_1a_2}
{\rm Erf}\Big[{a_1\over\sigma_1\sqrt2}\Big]
{\rm Erf}\Big[{a_2\over\sigma_2\sqrt2}\Big]\,.
\end{eqnarray}
It happens, however, that even in the presence of cross-correlations,
if the prior is wide ($a_i \geq 2\sigma_i$), then the
terms proportional to exponentials are negligible and the evidence
becomes, to very good approximation,
\begin{equation}\label{evidence2}
E(a_1,a_2) = {1\over4a_1a_2}\,
{\rm Erf}\Big[{a_2\over\sigma_2\sqrt{2(1-\rho^2)}}\Big]
{\rm Erf}\Big[{a_1\over\sigma_1\sqrt{2}}\Big] \,.
\end{equation}
Moreover, in that case, the error functions are very approximately
given by 1.

\vspace{5mm} {\bf $n$ variables}. Suppose we have $n$ correlated
variables, ${\bf x}=(x_1,\dots,x_n)$, with covariance matrix
\begin{equation}\label{corrn}
 C_n = \left(\begin{array}{cccc}
 C_{11}& C_{12}&\ldots& C_{1n}\\[2mm]
 C_{12}& C_{22}&\ldots& C_{2n}\\[2mm]
\vdots&\vdots&\ddots&\vdots\\[2mm]
 C_{1n}& C_{2n}&\ldots& C_{nn}
\end{array}\right) \,.
\end{equation}
In that case, the probability distribution function can be expressed as
\begin{equation}\label{fxn}
f({\bf x}) = {1\over(2\pi)^{n/2}\sqrt{\det C_n}} \exp\Big[\!
-\half {\bf x}^{_T} C_n^{-1}{\bf x}\Big]\,,
\end{equation}
which has the property that marginalizing over the last variable, $x_n$,
we obtain a correlated probability distribution function for the
$n-1$ variables, ${\bf x}=(x_1,\dots,x_{n-1})$,
\begin{equation}\label{fxn1}
f({\bf x}) = {1\over(2\pi)^{(n-1)/2}\sqrt{\det C_{n-1}}} \exp\Big[\!
-\half {\bf x}^{_T} C_{n-1}^{-1}{\bf x}\Big]\,,
\end{equation}
where the $ C_{n-1}$ covariance matrix is given by Eq.~(\ref{corrn})
without the last column and the last row.

We will now evaluate the evidence Eq.~(\ref{evx}) for this multivariate
Gaussian, starting with the integration over the last variable, $x_n$,
\begin{eqnarray}\nonumber
{1\over2a_n} \int_{-a_n}^{a_n} dx_n\ f({\bf x}) &\!=\!& 
{1\over(2\pi)^{(n-1)/2}\sqrt{\det C_{n-1}}} \exp\Big[\!
-\half {\bf x}^{_T} C_{n-1}^{-1}{\bf x}\Big] \\[2mm]\label{intn}
&&\times\left\{
{1\over2a_n}\,{\rm Erf}\left[{a_n\over\sqrt2}\sqrt{\det C_{n-1}\over
\det C_n}\right] + {\cal O}\Big(e^{-{a_n^2\,\det C_{n-1}
\over2\det C_n}}\Big)\right\}\,.
\end{eqnarray}
Integrating now over the next variable, $x_{n-1}$, we find
\begin{eqnarray}\nonumber
&&{1\over4a_na_{n-1}} \int_{-a_n}^{a_n} dx_n
\int_{-a_{n-1}}^{a_{n-1}} dx_{n-1}\ f({\bf x}) =
{1\over(2\pi)^{(n-2)/2}\sqrt{\det C_{n-2}}} \exp\Big[
-\half\,{\bf x}^{_T} C_{n-2}^{-1}{\bf x}\Big] \\[2mm]
\label{intn1} &&\times\left\{
{1\over4a_na_{n-1}}\,{\rm Erf}\left[{a_n\over\sqrt2}
\sqrt{\det C_{n-1}\over\det C_n}\right]\,
{\rm Erf}\left[{a_n\over\sqrt2}\sqrt{\det C_{n-2}\over
\det C_{n-1}}\right] + {\cal O}\Big(e^{-{a_n^2\,\det C_{n-1}
\over2\det C_n}}\Big)\right\}\,.
\end{eqnarray}
Continuing the integration over the priors, we end up with the
evidence for the $n$-dimensional distribution,
\begin{eqnarray}\nonumber
E(a_1,\dots,a_n) &\!=\!&  {1\over\prod_{p=1}^n 2a_p} 
\int_{-a_1}^{a_1}\!\dots
\int_{-a_n}^{a_n} d^n{\bf x}\ f({\bf x}) \\[2mm]\label{evidencen}
&\!=\!&\,\prod_{p=1}^n{1\over2a_p}{\rm Erf}\left[{a_p\over\sqrt2}
\sqrt{\det C_{p-1}\over\det C_p}\right] + {\cal O}
\left(\exp\Big[-\sum_{p=1}^n{a_p^2\,\det C_{p-1}
\over2\det C_p}\Big]\right)\,,
\end{eqnarray}
where the covariance matrices $ C_p$ are constructed as above,
by eliminating the $n-p$ last rows and columns, until we end up with
$ C_0\equiv1$. Note that the approximation is very good whenever
$\sum_{p=1}^n (a_p^2\,\det C_{p-1})/(2\det C_p)\gg1$, which is
often the case. Note also that we recover the previous result
Eq.~(\ref{evidence2}) for the particular case $n=2$.

In the limit that the cross-correlation between the $n$ variables 
vanishes, the evidence (\ref{evidencen}) reduces to the exact result
\begin{equation}\label{evidencexact}
E(a_1,\dots,a_n) = \prod_{p=1}^n{1\over2a_p}
{\rm Erf}\left[{a_p\over\sigma_p\sqrt2}\right]\,.
\end{equation}
Note that the evidence Eq.~(\ref{evidencen}) reflects correctly the limit 
in which we eliminate the need for a new variable $x_n$, by making its 
prior vanish, 
\begin{equation}
\lim_{a_n\to0}\ E(a_1,\dots,a_n)=E(a_1,\dots,a_{n-1})\,
{1\over\sqrt{2\pi}}\,\sqrt{\det C_{n-1}\over\det C_n}\,,
\end{equation}
and thus we recover in that limit a properly-normalized distribution,
$f(x_1,\dots,x_n)\to f(x_1,\dots,x_{n-1})$, while the inspection of
the likelihood function alone would not have been able to give a
reasonable answer.

On the other hand, in the case that our theoretical prejudice cannot
assign a concrete prior to a given variable, we see that the evidence
decreases as $1/2a$ as $a$ increases. Therefore, the Bayesian evidence
seems to be a very good discriminator between theoretical priors, and
penalizes including too many parameters, {\it a la} Occam's razor.

\subsection{Uncentered priors}

It is unlikely that the priors will actually be centred on the mean of the 
distribution, as the priors are not supposed to know what the data will tell us. 
We therefore need to generalize the above for uncentred priors. We continue to 
assume that the priors are top hats.

We also continue to assume for the moment that the
probability distribution is well approximated by a Gaussian with mean
value $\mu$. We will then use displaced variables $\tilde x_i = x_i -
\mu_i$, and write the Gaussian distribution function as in
Eq.~(\ref{fxn}). The normalized top-hat prior is now uncentered with
respect to the mean value,
\begin{equation}\label{abprior}
\pi(\tilde x;a,b) \equiv \left\{\begin{array}{ll}(a+b)^{-1}&\hspace{1cm}
-a < \tilde x < b\,,\\[2mm] 0 &\hspace{1cm} {\rm otherwise}\,.\end{array}
\right.
\end{equation}
For a single variable, the result is {\em exact},
\begin{equation}\label{evi1unc}
E(a;b) = \int_{-\infty}^{\infty} dx\,f(x)\,\pi(x;a,b) =
{1\over2a+2b}\left({\rm Erf}\left[{a\over\sigma\sqrt2}\right] + 
{\rm Erf}\left[{b\over\sigma\sqrt2}\right]\right)\,.
\end{equation}
where we are integrating over the displaced variable $\tilde x$, from
now on renamed as $x$. Note that we recover the result Eq.~(\ref{evidence1})
for the centered prior case in the limit $b\to a$.

For two variables, with distribution function Eq.~(\ref{f2}), the 
uncentered Bayesian evidence is
\begin{eqnarray}\label{evxab2}
E(a_1,a_2;b_1,b_2) &\!=\!& {1\over(a_1+b_1)(a_2+b_2)}
\int_{-a_1}^{b_1} dx_1\,\int_{-a_2}^{b_2} dx_2\,f(x_1,x_2) \\[2mm] 
&=& {1\over(2a_1+2b_1)(2a_2+2b_2)}\left\{
\left({\rm Erf}\left[{a_1\over\sigma_1\sqrt2}\right] + 
{\rm Erf}\left[{b_1\over\sigma_1\sqrt2}\right]\right)\right.
\\[2mm] \nonumber && \times 
\left({\rm Erf}\left[{a_2\over\sigma_2\sqrt{2(1-\rho^2)}}\right] + 
{\rm Erf}\left[{b_2\over\sigma_2\sqrt{2(1-\rho^2)}}\right]\right)
\\[2mm] \nonumber
&& \left.- \,{\rho\over2\pi\sqrt{1-\rho^2}}\left(
e^{-{a_1^2\over2\sigma_1^2}}-e^{-{b_1^2\over2\sigma_1^2}}\right)
\left(e^{-{a_2^2\over2\sigma_2^2(1-\rho^2)}} +
e^{-{b_2^2\over2\sigma_2^2(1-\rho^2)}}
\right)\right\}
\end{eqnarray}

The evidence for the multiple-variable case Eq.~(\ref{fxn}) is
\begin{equation}\label{evxab}
E({\bf a},{\bf b}) =
\int_{-\infty}^{\infty} d^n{\bf x}\,f({\bf x})\,
\prod_{i=1}^n \,\pi(x_i;a_i,b_i) = \prod_{i=1}^n (a_i+b_i)^{-1} \!
\int_{-a_1}^{b_1}d\tilde x_1\dots\int_{-a_n}^{b_n}
d\tilde x_n\,f(\tilde{\bf x})\,.
\end{equation}
Let us now evaluate it for the multivariate Gaussian Eq.~(\ref{fxn}), starting 
with
the integration over the last variable, $x_n$,
\begin{eqnarray}\nonumber
&&{1\over a_n+b_n} \int_{-a_n}^{b_n} dx_n\ f({\bf x}) \ = \ 
{1\over(2\pi)^{(n-1)/2}\sqrt{\det C_{n-1}}} \exp\Big[\!
-\half {\bf x}^{_T} C_{n-1}^{-1}{\bf x}\Big]\,
{1\over(2a_n+2b_n)} \\[2mm]
&&\times\left\{
{\rm Erf}\left[{a_n\over\sqrt2}\sqrt{\det C_{n-1}\over
\det C_n}\right] + 
{\rm Erf}\left[{b_n\over\sqrt2}\sqrt{\det C_{n-1}\over
\det C_n}\right] + {\cal O}\left(
e^{-{a_n^2\,\det C_{n-1}\over2\det C_n}} +
e^{-{b_n^2\,\det C_{n-1}\over2\det C_n}}\right)
\right\}\hspace{5mm}
\end{eqnarray}
Integrating now over the next variable, $x_{n-1}$, we find
\begin{eqnarray}\nonumber
&&{1\over(a_n+b_n)(a_{n-1}+b_{n-1})} \int_{-a_n}^{b_n} dx_n
\int_{-a_{n-1}}^{b_{n-1}} dx_{n-1}\ f({\bf x}) = \\[2mm]
&&{1\over(2\pi)^{(n-2)/2}\sqrt{\det C_{n-2}}} \exp\Big[
-\half\,{\bf x}^{_T} C_{n-2}^{-1}{\bf x}\Big]\,{1\over
(2a_n+2b_n)(2a_{n-1}+2b_{n-1})} \\[2mm]
&&\times\left\{\left({\rm Erf}\left[{a_n\over\sqrt2}
\sqrt{\det C_{n-1}\over\det C_n}\right]+
{\rm Erf}\left[{b_n\over\sqrt2}
\sqrt{\det C_{n-1}\over\det C_n}\right]\right)\right.\\[2mm]
&&\times\left({\rm Erf}\left[{a_{n-1}\over\sqrt2}\sqrt{\det C_{n-2}
\over\det C_{n-1}}\right]+
{\rm Erf}\left[{b_{n-1}\over\sqrt2}\sqrt{\det C_{n-2}\over
\det C_{n-1}}\right]\right)\\[2mm]\nonumber
&& + \ \left.{\cal O}\left(e^{-{a_n^2\,\det C_{n-1}
\over2\det C_n}} + e^{-{b_n^2\,\det C_{n-1}
\over2\det C_n}}\right)\times\left(e^{-{a_{n-1}^2\,\det C_{n-2}
\over2\det C_{n-1}}} + e^{-{b_{n-1}^2\,\det C_{n-2}
\over2\det C_{n-1}}}\right)\right\}\,.
\end{eqnarray}
Continuing the integration over the priors, we end up with the
evidence for the $n$-dimensional distribution,
\begin{eqnarray}\nonumber
E({\bf a},{\bf b}) &\!=\!&  {1\over\prod_{p=1}^n(a_p+b_p)} 
\int_{-a_1}^{b_1}\!\dots
\int_{-a_n}^{b_n} d^n{\bf x}\ f({\bf x}) \\[2mm]\label{evidencenab}
&\!=\!& \prod_{p=1}^n{1\over(2a_p+2b_p)}\left(
{\rm Erf}\left[{a_p\over\sqrt2}
\sqrt{\det C_{p-1}\over\det C_p}\right] +
{\rm Erf}\left[{b_p\over\sqrt2}
\sqrt{\det C_{p-1}\over\det C_p}\right]\right) \\[2mm]\nonumber
&& \hspace{2.5cm} + \
{\cal O}\left(\prod_{p=1}^n\left[\exp\Big(-{a_p^2\,\det C_{p-1}
\over2\det C_p}\Big)+\exp\Big(-{b_p^2\,\det C_{p-1}
\over2\det C_p}\Big)\right]\right)\,,
\end{eqnarray}
where the covariance matrices $ C_p$ are constructed as above, by
eliminating the $n-p$ last rows and columns, until $ C_0\equiv1$.  Note
that the approximation is very good whenever the exponents are large,
$\sum_{p=1}^n (a_p^2\,\det C_{p-1})/(2\det C_p)\gg1$, which is often the
case. Note also that we recover the expression of the evidence for the
centered priors Eq.~(\ref{evidencen}) in the limit $b\to a$.

Let us now evaluate the evidence for a distribution normalized to the
maximum of the likelihood distribution,
\begin{equation}
f({\bf x}) = {\cal L}_{\rm max}
\exp\Big[\!-\!\half {\bf x}^{_T} C_n^{-1}{\bf x}\Big]
\end{equation}
In this case, the evidence is given by Eq.~(\ref{evidencenab}), multiplied by a
factor ${\cal L}_{\rm max}\times(2\pi)^{n/2}\sqrt{\det C_n}$ from the
normalization. We can then evaluate the logarithm of the evidence,
ignoring the exponentially-small corrections, as
\begin{eqnarray}\nonumber
\ln E &=& \ln{\cal L}_{\rm max} + {n\over2}\ln(2\pi) + \half \ln \det C_n
- \sum_{p=1}^n\ln(2a_p+2b_p) \\[2mm]\label{evidencenabc}
&& \ + \sum_{p=1}^n \ln \left(
{\rm Erf}\left[{a_p\over\sqrt2}\sqrt{\det C_{p-1}\over \det C_p}\right]+
{\rm Erf}\left[{b_p\over\sqrt2}\sqrt{\det C_{p-1}\over \det C_p}\right]
\right)\,.
\end{eqnarray}

\vspace{5mm} {\bf Uncorrelated case}. Suppose we have a multivariate
Gaussian distribution without correlations between variables, i.e.
$ C_{ij}=\sigma_i^2\delta_{ij}$ is a diagonal matrix; then the evidence 
reads {\em exactly},
\begin{equation}\label{evidencenabuncor}
E({\bf a},{\bf b}) = {1\over\prod_{p=1}^n(a_p+b_p)} 
\int_{-a_1}^{b_1}\!\dots
\int_{-a_n}^{b_n} d^n{\bf x}\ f({\bf x}) = \prod_{p=1}^n{1\over2(a_p+b_p)}\left(
{\rm Erf}\left[{a_p\over\sigma_p\sqrt2}\right] +
{\rm Erf}\left[{b_p\over\sigma_p\sqrt2}\right]\right)\,,
\end{equation}
where $\sigma_p$ are the dispersions of each variable $\tilde x_p$, and thus
the logarithm of the evidence becomes
\begin{equation}\label{exactevn}
\ln E = \ln{\cal L}_{\rm max} + {n\over2}\ln(2\pi) + 
\sum_{p=1}^n \ln \sigma_p - \sum_{p=1}^n\ln(2a_p+2b_p) + 
\sum_{p=1}^n \ln \left(
{\rm Erf}\left[{a_p\over\sigma_p\sqrt2}\right] +
{\rm Erf}\left[{b_p\over\sigma_p\sqrt2}\right]\right)
\end{equation}

\vspace{5mm} 

{\bf Laplace approximation}.  The Laplacian approximation
to the evidence assumes the distribution is a correlated Gaussian, and
that the priors are large enough so that the whole distribution fits
easily inside them, in which case the error functions are approximately
unity and do not contribute to the evidence; from Eq.~(\ref{evidencenabc}) we 
now have
\begin{equation}\label{Laplace}
\ln E = \ln{\cal L}_{\rm max} + {n\over2}\ln(2\pi) + \half \ln \det C_n
- \sum_{p=1}^n\ln \Delta\theta_p\,,
\end{equation}
where $\Delta\theta_p=a_p+b_p$ is the parameter interval associated to
the prior.  In the next section we will compare the different
approximations.

\section{Non-Gaussian corrections}

The advantage of this method is that one can perform a systematic
computation of the evidence of a given model with its own priors,
given an arbitrary set of moments of the distribution. Here we will
consider the first two beyond the covariance matrix, i.e.~the
skewness and the kurtosis terms, see Eq.~(\ref{phi}).

\subsection{Skewness}

Let us start with the first correction to
the Gaussian approximation, the trilinear term $ B_{ijk}$. For
this, we write the generating functional (\ref{phi}) as
\begin{equation}\label{phi3}
\phi({\bf u}) = \exp\left[i\,\mu_i u_i - {1\over2!}\, C_{ij}
\,u_i u_j - \,{i\over3!}\, B_{ijk}\,u_i u_j u_k \right]\,.
\end{equation}
By performing a change of variable, $u_i = y_i - i\, C_{ik}^{-1}
(x_k - \mu_k)$, we can evaluate the Fourier transform integral and
obtain the properly-normalized probability distribution function
\begin{eqnarray}\nonumber
f({\bf x}) &\!=\!&{1\over(2\pi)^{n/2}\sqrt{\det C_n}}\exp\Big[\!
-\half {\bf x}^{_T} C_n^{-1}{\bf x}\Big]\\[2mm]\label{fx3}
&&\times \left(1 - \half  B_{ijk}\,
 C_{ij}^{-1} C_{kl}^{-1}\,x_l + {1\over6} B_{ijk}\,
 C_{il}^{-1} C_{jm}^{-1} C_{kn}^{-1}\,x_l x_m x_n\right)\,,
\end{eqnarray}
where $x_k$ are the displaced coordinates $(x_k - \mu_k)$.
This skewed distribution function satisfies
\begin{equation}
\langle x_i\rangle = 0\,, \hspace{1cm}
\langle x_i x_j\rangle =  C_{ij}\,, \hspace{1cm}
\langle x_i x_j x_k\rangle =  B_{ijk}\,, \hspace{1cm}
\langle x_i x_j x_k x_l\rangle = 0\,, 
\hspace{5mm}\dots
\end{equation}
as can be confirmed by direct evaluation.  Let us now compute the
evidence Eq.~(\ref{evx}) for this skewed model.  Since the extra terms in
the parenthesis of Eq.~(\ref{fx3}) are both odd functions of $x$, when
integrating over an even range like that of the centered top-hat prior
Eq.~(\ref{prior}), their contribution to the evidence vanish, and thus the
final evidence for the skewed model does not differ from that of the
Gaussian model Eq.~(\ref{evidencen}). In case the prior is off-centered
with respect to the mean, e.g. like in Eq.~(\ref{abprior}), then the
contribution of the odd terms to the evidence would not vanish.  Let
us evaluate their contribution.

For a single variable $(n=1)$, the correctly-normalized
likelihood function can be written as 
$$f(x) = {e^{-{x^2/2\sigma^2}}\over\sigma\sqrt{2\pi}}\,
\left(1 - {B\,x\over2\sigma^4} + {B\,x^3\over6\sigma^6}\right)\,,$$
satisfying $\langle x\rangle=0$, $\langle x^2\rangle=\sigma^2$, 
$\langle x^3\rangle=B$, and the Bayesian integral can be computed 
{\em exactly} as
\begin{equation}
E(a,b) = {1\over2a+2b}\left({\rm Erf}\left[{a\over\sigma\sqrt2}\right]+
{\rm Erf}\left[{b\over\sigma\sqrt2}\right]\right) - {B\sigma^{-3}\over
6\sqrt{2\pi}}\left[\Big(1-{a^2\over\sigma^2}\Big)\,
e^{-{a^2\over2\sigma^2}} -\Big(1-{b^2\over\sigma^2}\Big)\,
e^{-{b^2\over2\sigma^2}}\right]{1\over a+b}\,.
\end{equation}
Note that for even (centered) priors, with $b=a$, the evidence reduces
to Eq.~(\ref{evidence1}).

For an arbitrary number of variables, the computation is more
complicated.  Let us start with the $n$-th variable and, in order to
compute the integral, let us define the auxiliary function
\begin{eqnarray}\nonumber
g(\lambda) &\!=\!& \int_{-a_n}^{b_n} dx_n\,x_n\,
{\exp\Big[\!-\!{\lambda\over2}{\bf x}^{_T} C_n^{-1}{\bf x}\Big]
\over (2\pi)^{n/2}\sqrt{\det C_n}} \ = \ 
{\exp\Big[\!-\!\half {\bf x}^{_T} C_{n-1}^{-1}
{\bf x}\Big]\over(2\pi)^{(n-1)/2}\sqrt{\det C_{n-1}}} \times \\[2mm]
&&\times {1\over\lambda\sqrt{2\pi}}\,\left(\exp\Big[\!-\!
{\lambda a_n^2\over2}{\det C_{n-1}\over\det C_n}\Big] -
\exp\Big[\!-\!{\lambda b_n^2\over2}{\det C_{n-1}\over\det C_n}\Big]
\right)\,,
\end{eqnarray}
such that, using ${\rm Erf}'[x] = {2\over\sqrt\pi}\,e^{-x^2}$,
\begin{eqnarray}\nonumber
&&-2g'(\lambda=1) = \int_{-a_n}^{b_n} dx_n\,x_n\,
{({\bf x}^{_T} C_n^{-1}{\bf x})\,\exp\Big[\!-\!{1\over2}
{\bf x}^{_T} C_n^{-1}{\bf x}\Big]\over(2\pi)^{n/2}\sqrt{\det C_n}} 
\ = \ {\exp\Big[\!-\!\half {\bf x}^{_T} C_{n-1}^{-1}{\bf x}\Big]
\over(2\pi)^{(n-1)/2}\sqrt{\det C_{n-1}}} \times \\[2mm]\label{2gp}
&&\hspace{1cm}\times {1\over\sqrt{2\pi}}\,\left\{
\left(2+a_n^2{\det C_{n-1}\over\det C_n}\right)
\exp\Big[\!-\!{a_n^2\over2}{\det C_{n-1}\over\det C_n}\Big] -
\left(2+b_n^2{\det C_{n-1}\over\det C_n}\right)
\exp\Big[\!-\!{b_n^2\over2}{\det C_{n-1}\over\det C_n}\Big]
\right\}\,.
\end{eqnarray}
Therefore, with the use of Eq.~(\ref{2gp}), the integral of the
skewness-corrected distribution function Eq.~(\ref{fx3}) over the $x_n$
uncentered prior, becomes
\begin{eqnarray}\nonumber
&&\int_{-a_n}^{b_n} dx_n\ f({\bf x}) =
{\exp\Big[\!-\!\half {\bf x}^{_T} C_{n-1}^{-1}{\bf x}\Big]\over
(2\pi)^{(n-1)/2}\sqrt{\det C_{n-1}}}\,\left\{{1\over2}\,
\left({\rm Erf}\left[{a_n\over\sqrt2}
\sqrt{\det C_{n-1}\over\det C_n}\right] + {\rm Erf}\left[{b_n\over\sqrt2}
\sqrt{\det C_{n-1}\over\det C_n}\right]\right)\right.\\[2mm]
&&- \left. {1\over6} B_{ijn}\,C_{ij}^{-1}{1\over\sqrt{2\pi}}
\sqrt{\det C_{n-1}\over\det C_n}
\left[\left(1-a_n^2{\det C_{n-1}\over\det C_n}\right)
e^{-{a_n^2\,\det C_{n-1}\over2\det C_n}} -
\left(1-b_n^2{\det C_{n-1}\over\det C_n}\right)
e^{-{b_n^2\,\det C_{n-1}\over2\det C_n}}
\right]\right\}\,.\label{int3}
\end{eqnarray}
Let us define two new functions,
\begin{eqnarray}\label{EF}
E_i(a_i,b_i) &\!=\!& {1\over2}\,\left({\rm Erf}\left[{a_i\over\sqrt2}
\sqrt{\det C_{i-1}\over\det C_i}\right] + {\rm Erf}\left[{b_i\over\sqrt2}
\sqrt{\det C_{i-1}\over\det C_i}\right]\right)\,,\\[2mm]\nonumber
F_i(a_i,b_i) &\!=\!& {1\over6\sqrt{2\pi}}\sqrt{\det C_{i-1}\over\det C_i}
\left[\left(1-a_i^2{\det C_{i-1}\over\det C_i}\right)
e^{-{a_i^2\,\det C_{i-1}\over2\det C_i}} -
\left(1-b_i^2{\det C_{i-1}\over\det C_i}\right)
e^{-{b_i^2\,\det C_{i-1}\over2\det C_i}}\right]\,.
\end{eqnarray}
Integrating iteratively over $x_{n-1},\dots,x_1$, we end up with the
Bayesian evidence for the third-order-corrected probability distribution
function $f({\bf x})$,
\begin{equation}\label{eviskew}
E({\bf a},{\bf b}) = \prod_{p=1}^n\, {E_p(a_p,b_p)\over(a_p+b_p)}\,
\left[1 - \sum_{k=1}^n\,B_{ijk}\,C_{ij}^{-1}\,{F_k(a_k,b_k)\over E_k(a_k,b_k)}
\right]\,.
\end{equation}
Unless $ B_{ijk}\, C_{ij}^{-1}$ is very large, the correction to the
error function is exponentially suppressed, and we do not expect
significant departures from the Gaussian case
Eq.~(\ref{evidencen}). Note also that if the prior is symmetric, it is
easy to see that the skewness part of the integral vanishes,
$F_k(a_k,b_k)\to 0$, as can be checked explicitly by taking $b_k\to a_k$.

\subsection{Kurtosis}

The next correction beyond skewness is
the fourth order moment or kurtosis, given by the $ D_{ijkl}$
term in Eq.~(\ref{phi}). Let us ignore for the moment the third order
skewness and write
\begin{equation}\label{phi4}
\phi({\bf u}) = \exp\left[i\,\mu_i u_i - {1\over2!}\, C_{ij}
\,u_i u_j + \,{1\over4!}\, D_{ijkl}\,u_i u_j u_k u_l\right]\,.
\end{equation}
By performing the same change of variables, $u_i = y_i - i\,
C_{ik}^{-1} (x_k - \mu_k)$, we can now compute the Fourier transform
and obtain the properly-normalized probability distribution function
\begin{eqnarray}\nonumber
f({\bf x}) &\!=\!&{1\over(2\pi)^{n/2}\sqrt{\det C_n}}\exp\Big[\!
-\half {\bf x}^{_T} C_n^{-1}{\bf x}\Big]
\left(1 + {1\over8}  D_{ijkl}\,
 C_{ij}^{-1} C_{kl}^{-1} \right. \\[2mm]\label{fx4}
&& \left. - {1\over4}  D_{ijkl}\,
 C_{ij}^{-1} C_{km}^{-1} C_{ln}^{-1}\,x_m x_n +
{1\over24}  D_{ijkl}\,
 C_{im}^{-1} C_{jn}^{-1} C_{kp}^{-1} C_{lq}^{-1}
\,x_m x_n x_p x_q \right)\,.
\end{eqnarray}
Performing the integrals, it is easy to see that this distribution
satisfies
\begin{equation}
\langle x_i x_j\rangle =  C_{ij}\,, \hspace{1cm}
\langle x_i x_j x_k x_l\rangle =  D_{ijkl} + 
 C_{ij} C_{kl} +  C_{ik} C_{jl} + 
 C_{il} C_{jk}\,, \hspace{5mm}\dots
\end{equation}
Note that in order for the new likelihood distribution (\ref{fx4}) to be
positive definite, it is required that $D_{ijkl} C_{ij}^{-1} C_{kl}^{-1}<4$,
and if we impose that there is only one maximum at the center, then it
must satisfy $D_{ijkl} C_{ij}^{-1} C_{kl}^{-1}<2$. These conditions impose
bounds on the maximum possible deviation of the evidence from a that of a
gaussian.

Let us now compute the evidence Eq.~(\ref{evx}) for this kurtosis model.
The extra terms in the parenthesis of Eq.~(\ref{fx4}) are both
even functions of $x$, and we cannot ignore them, even for 
centered priors.

For a single variable $(n=1)$, the correctly-normalized
likelihood function can be written as 
$$f(x) = {e^{-{x^2\over2\sigma^2}}\over\sigma\sqrt{2\pi}}\, \left(1 +
{D\over8\sigma^4} - {D\,x^2\over4\sigma^6} +
{D\,x^4\over24\sigma^8}\right)\,,$$ satisfying $\langle x\rangle=0$,
$\langle x^2\rangle=\sigma^2$, $\langle x^3\rangle=0$, $\langle
x^4\rangle=D+3\sigma^4$, etc. The Bayesian integral can be computed
{\em exactly} as
\begin{equation}
E(a,b) = {1\over2a+2b}\left({\rm Erf}\left[{a\over\sigma\sqrt2}\right]+
{\rm Erf}\left[{b\over\sigma\sqrt2}\right]\right) + {D\sigma^{-4}\over
8\sqrt{2\pi}}\left({a\over\sigma}\Big(1-{a^2\over3\sigma^2}\Big)\,
e^{-{a^2\over2\sigma^2}} + {b\over\sigma}\Big(1-{b^2\over3\sigma^2}\Big)\,
e^{-{b^2\over2\sigma^2}}\right){1\over a+b}\,.
\end{equation}

For arbitrary number of variables, the computation is again much more
complicated.  Let us start with the $n$-th variable and, in order to
compute the first integral, let us define a new auxiliary function
\begin{eqnarray}\nonumber
h(\lambda) &\!=\!& \int_{-a_n}^{b_n} dx_n\,{\exp\Big[\!-\!
{\lambda\over2}{\bf x}^{_T} C_n^{-1}{\bf x}\Big]\over (2\pi)^{n/2}
\sqrt{\det C_n}} \ = \ {\exp\Big[\!-\!\half {\bf x}^{_T} C_{n-1}^{-1}
{\bf x}\Big]\over(2\pi)^{(n-1)/2}\sqrt{\det C_{n-1}}} \times \\[2mm]
&&\times {1\over2\sqrt\lambda}\,\left(
{\rm Erf}\left[{a_n\sqrt\lambda\over\sqrt2}
\sqrt{\det C_{n-1}\over\det C_n}\right] +
{\rm Erf}\left[{b_n\sqrt\lambda\over\sqrt2}
\sqrt{\det C_{n-1}\over\det C_n}\right]\right)\,,
\end{eqnarray}
such that, 
\begin{eqnarray}\nonumber
-2h'(\lambda=1) &\!=\!& \int_{-a_n}^{b_n} dx_n\,
{({\bf x}^{_T} C_n^{-1}{\bf x})\,\exp\Big[\!-\!{1\over2}
{\bf x}^{_T} C_n^{-1}{\bf x}\Big]\over(2\pi)^{n/2}\sqrt{\det C_n}} 
\ = \ {\exp\Big[\!-\!\half {\bf x}^{_T} C_{n-1}^{-1}{\bf x}\Big]
\over(2\pi)^{(n-1)/2}\sqrt{\det C_{n-1}}} \times \\[2mm]\label{2hp}
&&\times \left\{\half\,\left({\rm Erf}\left[{a_n\over\sqrt2}
\sqrt{\det C_{n-1}\over\det C_n}\right] + {\rm Erf}\left[{b_n\over\sqrt2}
\sqrt{\det C_{n-1}\over\det C_n}\right]\right)\right. \\[2mm]\nonumber
&& -\ \left.{1\over\sqrt{2\pi}}\sqrt{\det C_{n-1}\over\det C_n}\,
\left(a_n\,\exp\Big[\!-\!{a_n^2\over2}{\det C_{n-1}\over\det C_n}\Big] +
b_n\,\exp\Big[\!-\!{b_n^2\over2}{\det C_{n-1}\over\det C_n}\Big]\right)
\right\}\,.
\end{eqnarray}
\begin{eqnarray}\nonumber
4h''(\lambda=1) &\!=\!& \int_{-a_n}^{b_n} dx_n\,
{({\bf x}^{_T} C_n^{-1}{\bf x})^2\,\exp\Big[\!-\!{1\over2}
{\bf x}^{_T} C_n^{-1}{\bf x}\Big]\over(2\pi)^n\sqrt{\det C_n}}
\ = \ {\exp\Big[\!-\!\half {\bf x}^{_T} C_{n-1}^{-1}{\bf x}\Big]
\over(2\pi)^{(n-1)/2}\sqrt{\det C_{n-1}}} \times \\[2mm]\label{4hpp}
&&\times \left\{
{3\over2}\,\left({\rm Erf}\left[{a_n\over\sqrt2}
\sqrt{\det C_{n-1}\over\det C_n}\right] + {\rm Erf}\left[{b_n\over\sqrt2}
\sqrt{\det C_{n-1}\over\det C_n}\right]\right)\right. \\[2mm]\nonumber
&& -\ {3\over\sqrt{2\pi}}\sqrt{\det C_{n-1}\over\det C_n}\,
\left(a_n\,\exp\Big[\!-\!{a_n^2\over2}{\det C_{n-1}\over\det C_n}\Big] +
b_n\,\exp\Big[\!-\!{b_n^2\over2}{\det C_{n-1}\over\det C_n}\Big]\right)
\\[2mm]\nonumber
&& -\ \left.
{a_n^2\over\sqrt{2\pi}}\left({\det C_{n-1}\over\det C_n}\right)^{3/2}
\left(a_n\,\exp\Big[\!-\!{a_n^2\over2}{\det C_{n-1}\over\det C_n}\Big] +
b_n\,\exp\Big[\!-\!{b_n^2\over2}{\det C_{n-1}\over\det C_n}\Big]\right)
\right\}\,.
\end{eqnarray}
Therefore, with the use of Eqs.~(\ref{2hp}) and (\ref{4hpp}), the
integral of the kurtosis-corrected distribution function (\ref{fx4})
over the $x_n$ prior, becomes
\begin{eqnarray}
&&\int_{-a_n}^{b_n} dx_n\ f({\bf x}) =
{\exp\Big[\!-\!\half {\bf x}^{_T} C_{n-1}^{-1}{\bf x}\Big]\over
(2\pi)^{(n-1)/2}\sqrt{\det C_{n-1}}}
\left\{\half\,\left({\rm Erf}\left[{a_n\over\sqrt2}
\sqrt{\det C_{n-1}\over\det C_n}\right] + {\rm Erf}\left[{b_n\over\sqrt2}
\sqrt{\det C_{n-1}\over\det C_n}\right]\right)\right.+\\[2mm]\nonumber
&&+\left. {1\over8} D_{ijkl}\,C_{ij}^{-1} C_{kl}^{-1}{1\over\sqrt{2\pi}}
\sqrt{\det C_{n-1}\over\det C_n}\left[
a_n\!\left(1-{a_n^2\over3}{\det C_{n-1}\over\det C_n}\right)
e^{-{a_n^2\,\det C_{n-1}\over2\det C_n}} +
b_n\!\left(1-{b_n^2\over3}{\det C_{n-1}\over\det C_n}\right)
e^{-{b_n^2\,\det C_{n-1}\over2\det C_n}}
\right]\right\}\,.\label{int4}
\end{eqnarray}
We can now define a new function
\begin{equation}\label{G}
G_i(a_i,b_i) = {1\over8\sqrt{2\pi}}\sqrt{\det C_{i-1}\over\det C_i}
\left[a_i\left(1-{a_i^2\over3}{\det C_{i-1}\over\det C_i}\right)
e^{-{a_i^2\,\det C_{i-1}\over2\det C_i}} -
b_i\left(1-{b_i^2\over3}{\det C_{i-1}\over\det C_i}\right)
e^{-{b_i^2\,\det C_{i-1}\over2\det C_i}}\right]\,.
\end{equation}
Integrating iteratively over $x_{n-1},\dots,x_1$, we end up with the
Bayesian evidence for the fourth-order-corrected probability distribution
function $f({\bf x})$,
\begin{equation}\label{evikurt}
E({\bf a},{\bf b}) = \prod_{p=1}^n\, {E_p(a_p,b_p)\over(a_p+b_p)}\,
\left[1 + D_{ijkl}\,C_{ij}^{-1}\,C_{kl}^{-1}\,\sum_{m=1}^n\,
{G_m(a_m,b_m)\over E_m(a_m,b_m)}\right]\,.
\end{equation}
so, unless $ D_{ijkl}\, C_{ij}^{-1} C_{kl}^{-1}$ is
very large, the correction to the error function is exponentially
suppressed, and we do not expect significant departures from the
Gaussian case, Eq.~(\ref{evidencen}).

In order to compare models it is customary to compute the logarithm of
the evidence. Let us assume that we are given a
likelihood distribution function normalized by the maximum likelihood,
and with corrections up to fourth order,
\begin{eqnarray}\nonumber
&&f({\bf x}) = {\cal L}_{\rm max}
\exp\Big[\!-\!\half {\bf x}^{_T} C_n^{-1}{\bf x}\Big]
\left(1 + {1\over8}  D_{ijkl}\, C_{ij}^{-1} C_{kl}^{-1}\right)^{-1}
\!\!\left(1 - \half  B_{ijk}\, C_{ij}^{-1} C_{kl}^{-1}\,x_l 
+ {1\over6} B_{ijk}\, C_{il}^{-1} C_{jm}^{-1}
 C_{kn}^{-1}\,x_l x_m x_n \right. \\[2mm]\label{fx34}
&& \hspace{1cm} \left. +\ {1\over8}  D_{ijkl}\, C_{ij}^{-1} C_{kl}^{-1}
- {1\over4}  D_{ijkl}\, C_{ij}^{-1} C_{km}^{-1}
 C_{ln}^{-1}\,x_m x_n + {1\over24}  D_{ijkl}\,
 C_{im}^{-1} C_{jn}^{-1} C_{kp}^{-1} C_{lq}^{-1}
\,x_m x_n x_p x_q \right)\,.
\end{eqnarray}
Note that it is normalized so that the maximum corresponds to the
mean-centered distribution, i.e. ${\bf x}=0$.  In this case, the
evidence of the normalized distribution is given by
\begin{eqnarray}\label{eviskewkurt}
&&E({\bf a},{\bf b}) = {\cal L}_{\rm max}\ (2\pi)^{n/2}\sqrt{\det C_n}
\left(1 + {1\over8}  D_{ijkl}\, C_{ij}^{-1} C_{kl}^{-1}\right)^{-1}\times 
\\[2mm]\nonumber
&&\hspace{2cm} \prod_{p=1}^n\, {E_p(a_p,b_p)\over(a_p+b_p)}\,
\left[1 - \sum_{k=1}^n\,B_{ijk}\,C_{ij}^{-1}\,{F_k(a_k,b_k)\over E_k(a_k,b_k)}
+ D_{ijkl}\,C_{ij}^{-1}\,C_{kl}^{-1}\,\sum_{m=1}^n\,
{G_m(a_m,b_m)\over E_m(a_m,b_m)}\right]\,.
\end{eqnarray}
We can then evaluate the logarithm of the evidence by
\begin{eqnarray}\nonumber
\ln E &\!=\!& \ln{\cal L}_{\rm max} + {n\over2}\ln(2\pi) + \half \ln \det C_n 
- \ln\left(1 + {1\over8}  D_{ijkl}\, C_{ij}^{-1} C_{kl}^{-1}\right)
- \sum_{p=1}^n\ln(2a_p+2b_p) \\[2mm]\label{eviuncen}
&&+\ \sum_{p=1}^n \ln \left(
{\rm Erf}\left[{a_p\over\sqrt2}\sqrt{\det C_{p-1}\over \det C_p}\right]+
{\rm Erf}\left[{b_p\over\sqrt2}\sqrt{\det C_{p-1}\over \det C_p}\right]
\right)\\[2mm]\nonumber
&&+\ \ln\left(1 - \sum_{k=1}^n\,B_{ijk}\,C_{ij}^{-1}\,{F_k(a_k,b_k)\over 
E_k(a_k,b_k)} + D_{ijkl}\,C_{ij}^{-1}\,C_{kl}^{-1}\,\sum_{m=1}^n\,
{G_m(a_m,b_m)\over E_m(a_m,b_m)}
\right)\,. \label{eq:ng}
\end{eqnarray}
Note that the condition $D_{ijkl} C_{ij}^{-1} C_{kl}^{-1}<2$ constrains the 
maximum amount that the kurtosis corrections can contribute to the evidence.

\vspace{5mm} {\bf Uncorrelated case}. 
In the case where the likelihood distribution had no correlations among the
different variables, the {\em exact} expression for the Bayesian
evidence is
\begin{eqnarray}\label{exact}
&&\ln E = \ln{\cal L}_{\rm max} + {n\over2}\ln(2\pi) + \sum_{p=1}^n 
\ln \sigma_p - \sum_{p=1}^n\ln(2a_p+2b_p) + \sum_{p=1}^n \ln \left( 
{\rm Erf}\left[{a_p\over\sigma_p\sqrt2}\right] +
{\rm Erf}\left[{b_p\over\sigma_p\sqrt2}\right]\right)\\[2mm]\nonumber
&&\hspace{5mm}-\ \ln\left(1 + {1\over8} 
D_{iijj}\,\sigma_i^{-2}\sigma_j^{-2}\right)
+\ \ln\left(1 - \sum_{k=1}^n\,B_{iik}\,\sigma_k^{-2}\,{F_k(a_k,b_k)\over 
E_k(a_k,b_k)} + D_{iijj}\,\sigma_i^{-2}\sigma_j^{-2}\,\sum_{m=1}^n\,
{G_m(a_m,b_m)\over E_m(a_m,b_m)}
\right)\,,
\end{eqnarray}
where $\sigma_p$ are the corresponding dispersions of variables $x_p$,
and the functions $E_i, F_i$ and $G_i$ are the corresponding limiting
functions of Eqs.~(\ref{EF}) and (\ref{G}) for uncorrelated matrices.

\section{Model comparison}

Finally we turn to specific applications of the formalism discussed above. 
Initially 
we will carry out some toy model tests of its performance, and then examine real 
cosmological applications for which we previously obtained results by 
thermodynamic integration \cite{Beltran2005}.

\begin{figure*}[ht]
\includegraphics[width=6cm,angle=-90]{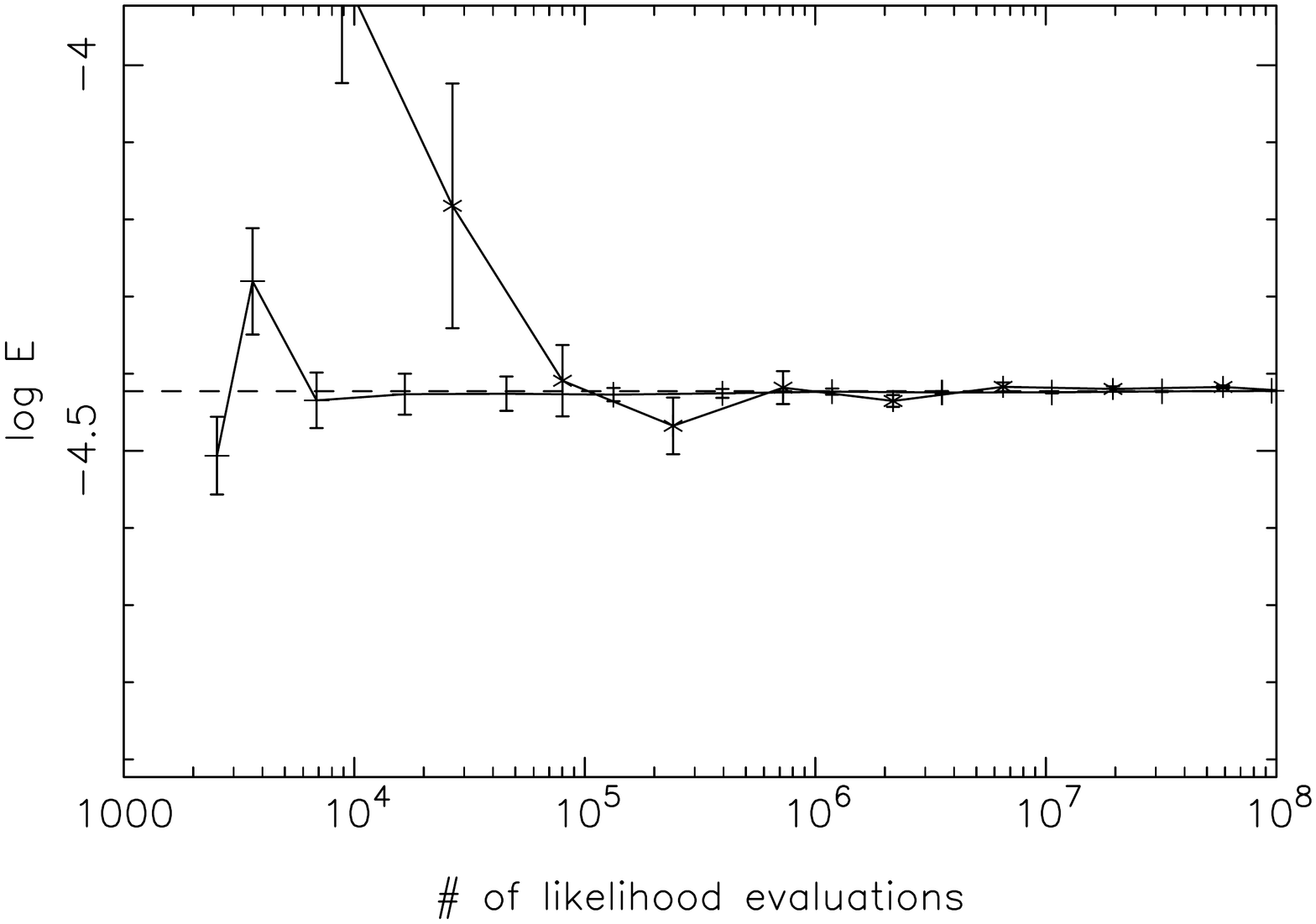} 
\includegraphics[width=6cm,angle=-90]{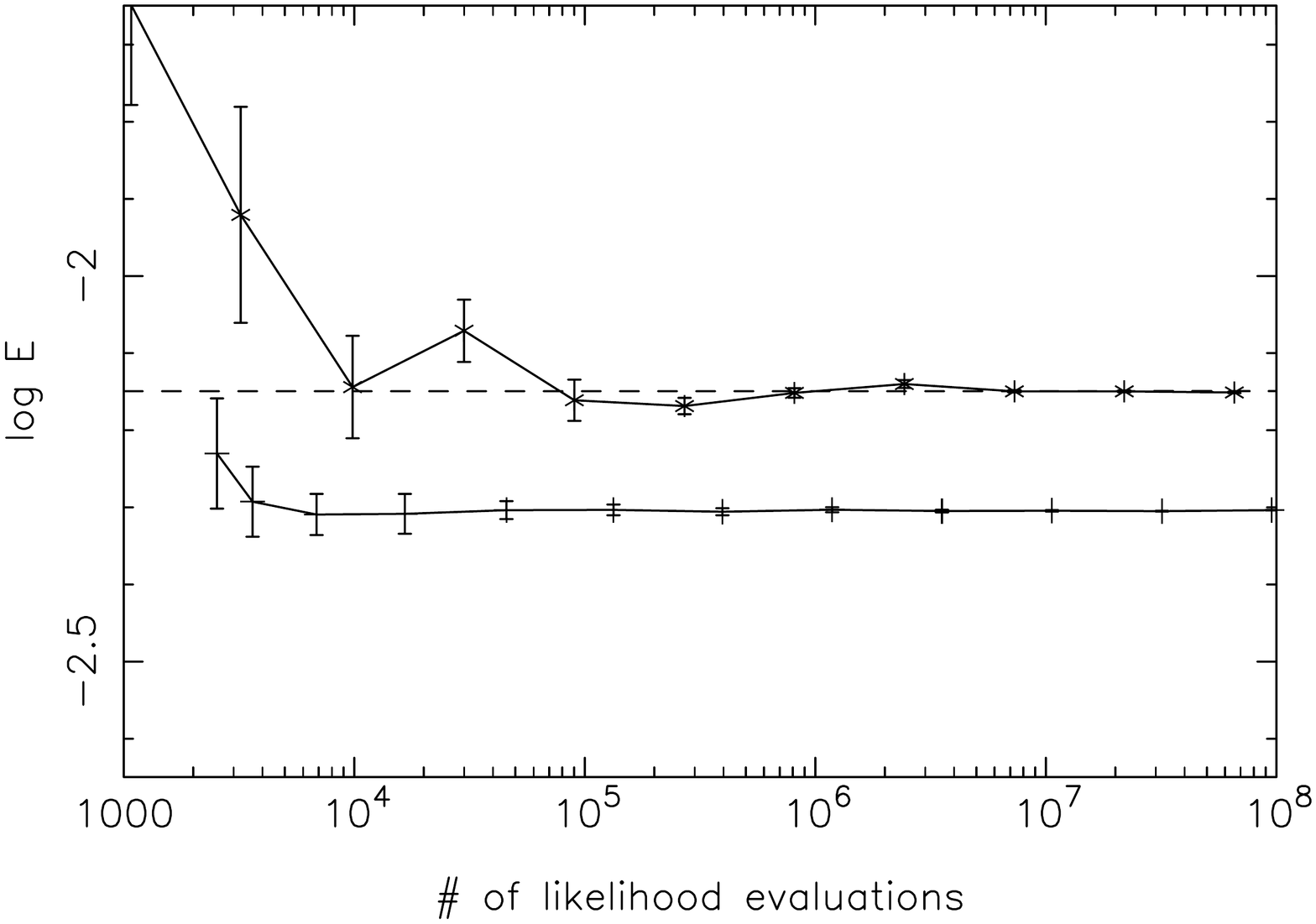}\\ 
\includegraphics[width=6cm,angle=-90]{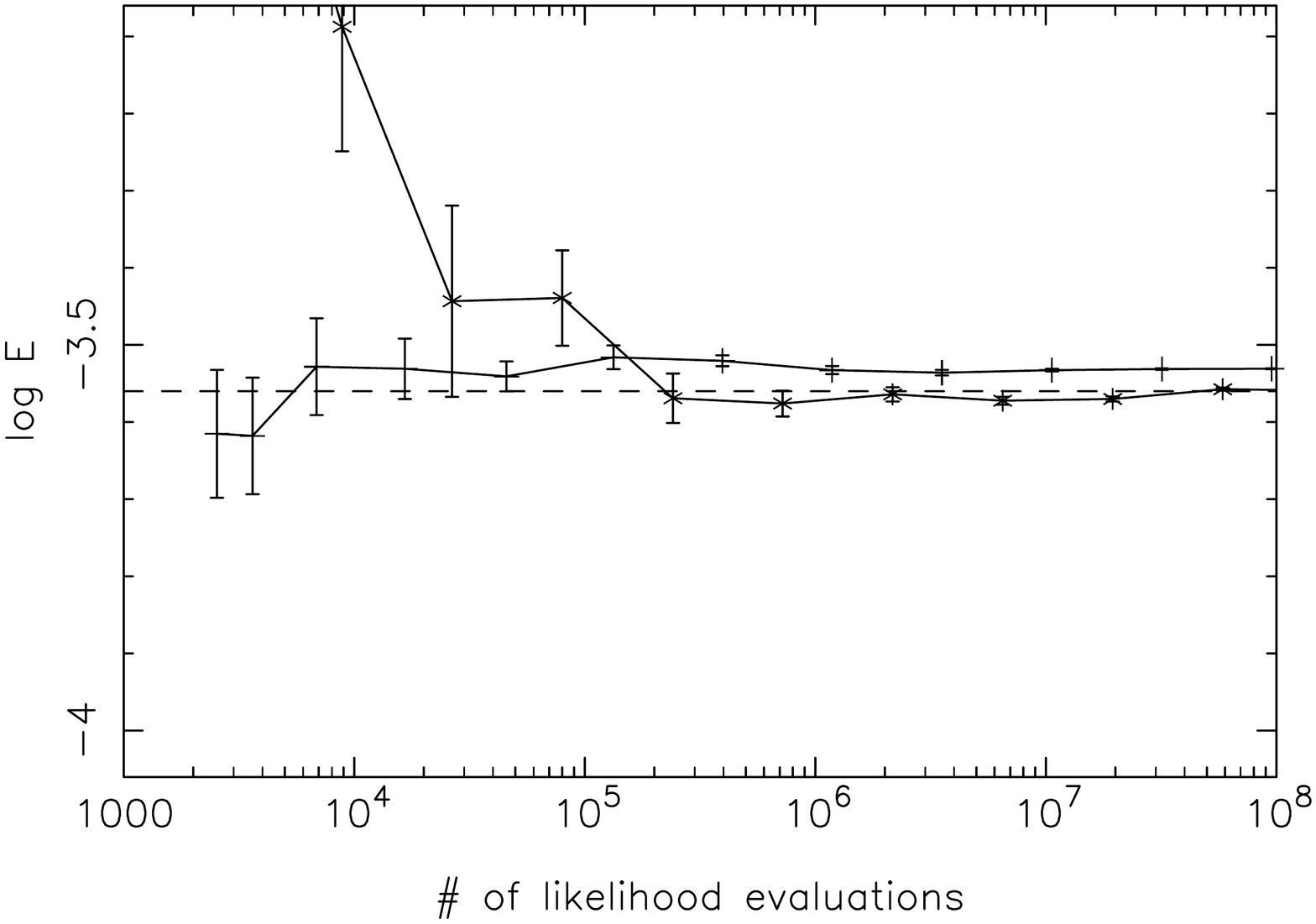} 
\includegraphics[width=6cm,angle=-90]{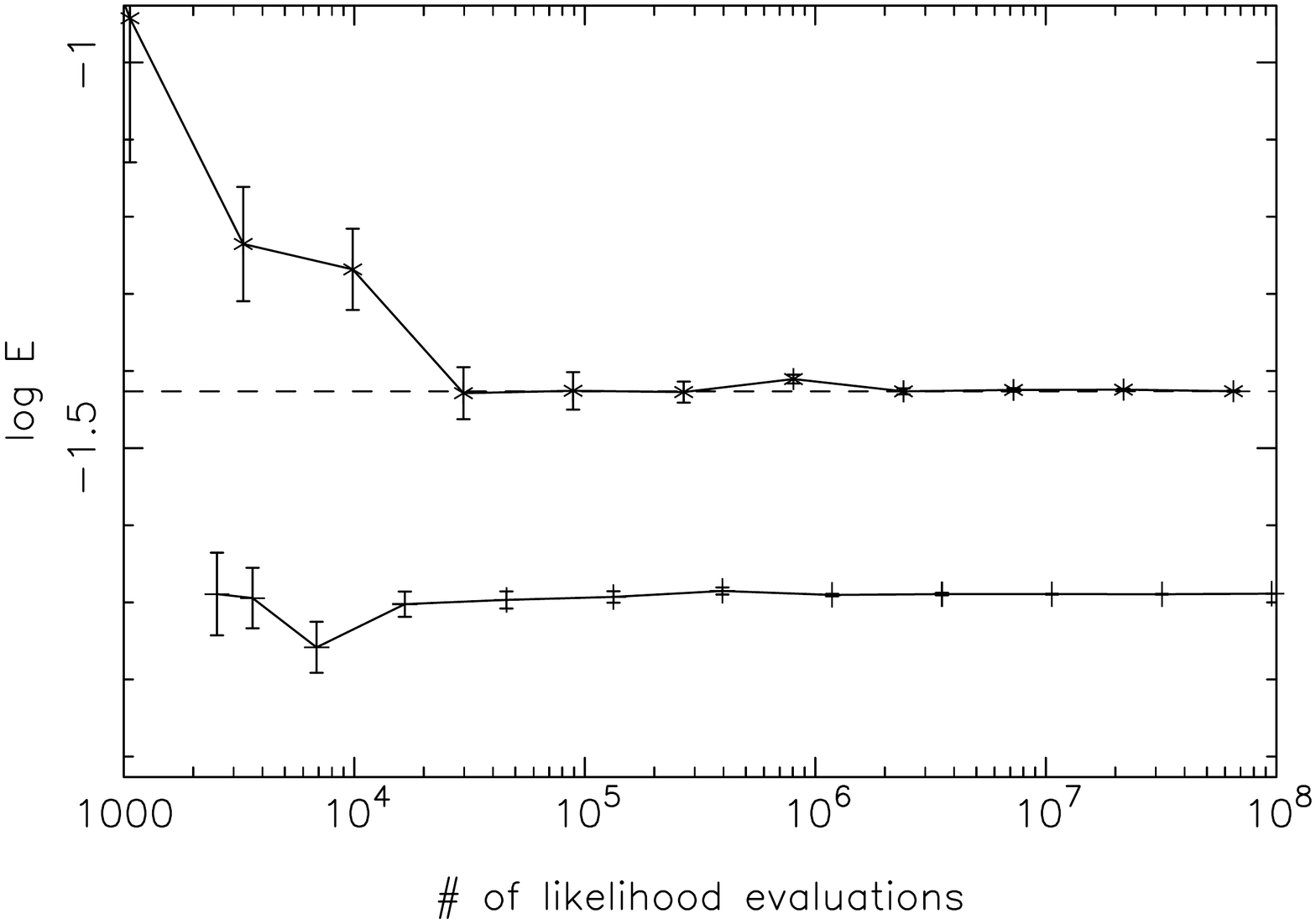}
\caption{This figure shows the calculated evidence as a function of  the number of likelihood evaluations.  Note that the horizontal axis  is logarithmic. The solid line corresponds to the thermodynamic  integration. The dotted line and dot-dashed lines are the analytical  methods with and without non-Gaussian corrections applied.  The  horizontal dashed line is the number obtained by the direct  integration. The upper two panels correspond to ${\cal L}_{g}$,  while the lower two to ${\cal L}_{ng}$. The left-hand side panels  correspond to wide flat priors of $(-7,10)$ on both parameters,  while the right-hand side to the narrow priors of $(-2,3)$ on both  parameters. See text for discussion. \label{fig:noevals}}
\end{figure*}

\subsection{A baby-toy model comparison}

We begin with a very simple two-dimensional toy model.  The purpose of
this section is to illustrate the ineffectiveness of the thermodynamic
integration and to give an indication of the performance of the method
we propose here. In addition, the two-dimensional model is simple
enough to allow a brute-force direct numerical integration of evidence
allowing us to check the accuracy at the same time.  We use the
following two forms of likelihood:
\begin{eqnarray}
{\cal L}_{g} (x,y) &=& \exp \left[-\frac{2x^2-2(y-1)^2-xy}{2}\right]\\
{\cal L}_{ng} (x,y) &=& \exp
\left[-\frac{2x^2-2(y-1)^2-xy}{2}\right]+
\exp \left[-\frac{2x^2 - 2y^2  - 3xy}{2} \right]
\end{eqnarray}
The subscripts $g$ and $ng$ indicate the Gaussian and non-Gaussian
cases respectively. 

Firstly, we calculate the evidence by the analytical method using
Eqs.~(\ref{exactevn}) and (\ref{eq:ng}) and covariance matrices
inferred from sampling the likelihood using the vanilla
Metropolis--Hastings algorithm with fixed proposal widths. Chains
ranging from few to several million samples were used. We also
calculate evidence using thermodynamic algorithm explained in
Ref.~\cite{Beltran2005}. Again, we vary algorithm parameters to get
evidence values of varying accuracy. The resulting evidence as a
function of number of likelihood evaluations is plotted in the Figure
\ref{fig:noevals}, together with the correct value inferred by direct
numerical integration. The number of likelihood evaluations is crucial
as this is the time-limiting step in the cosmological parameter
estimation and model comparison exercises.  The results are what could
have been anticipated.  We note that the size of the prior does not
seem to be of crucial importance. This is comforting, given that the
analytical method requires the knowledge of the \emph{true} covariance
information, while we can only supply a covariance matrix estimated
from the prior-truncated likelihood. We also note that the
thermodynamic integration converges to the correct value in all
cases. However, it does so after very many likelihood evaluations;
typically about a million or so even for a two-dimensional problem.
The analytical method becomes limited by systematics already by the
ten-thousand samples. For Gaussian case, there is no systematic by
construction, while the non-gaussian case suffers a systematic of
about $0.1$ in $\ln E$. The non-Gaussian correction reduces the error
by about a half and thus correctly estimates the uncertainty associated
with the purely Gaussian approximation. In the case of wide priors, the
only non-Gaussian correction of an appreciable size is the
$\ln(1+D_{ijkl}C^{-1}_{ij}C^{-1}_{kl}/8)$.

\subsection{A toy model comparison}

We now proceed by calculating the Bayesian evidence for simple toy models with 5
and 6 parameters, shown in Table~I. The purpose is to 
compare results with those obtained  from thermodynamic integration
again, but this time using a model that bears more resemblance to 
a typical problem one encounters in cosmology.

\begin{table}[thb]
\begin{center}
\begin{tabular}{c c c c}
\hline Parameter &\hspace{5mm} Mean&\hspace{10mm}Prior Range\hspace{5mm}
&\hspace{5mm}Model\hspace{3mm}\\
\hline
$x_1$ &\hspace{5mm}0.022&\hspace{5mm}[0.0001,\,0.044]&\hspace{5mm}toy5,toy6\\
$x_2$ &\hspace{5mm}0.12&\hspace{5mm}[0.001,\,0.3]&\hspace{5mm}toy5,toy6\\
$x_3$ &\hspace{5mm}1.04&\hspace{5mm}[0.8,\,1.4]&\hspace{5mm}toy5,toy6 \\
$x_4$ &\hspace{5mm}0.1&\hspace{5mm}[0.01,\,0.3]&\hspace{5mm}toy5,toy6 \\
$x_5$ &\hspace{5mm}3.1&\hspace{5mm}[2.6,\,3.6]&\hspace{5mm}toy5,toy6 \\
$x_6$ &\hspace{5mm}0.98&\hspace{5mm}[0.5,\,1.5]&\hspace{5mm}toy6 \\
\hline
\end{tabular}
\end{center}
\caption{\label{params2} The parameters used in the analytical
evaluation of the toy model evidences, with 5 and 6 parameters
respectively. The maximum likelihod of the toy models is taken
(arbitrarily) to be ${\cal L}_{\rm max} = 1$.}
\end{table}

 Beginning with the five-parameter model, we assume first that it
has an uncorrelated multivariate Gaussian likelihood
distribution. In this case the aim is to  test the thermodynamic
integration method, which gives $\ln E_{\rm toy5}^{\rm  num} =
-8.65\pm0.03$, while the exact expression gives $\ln E_{\rm
toy5}^{\rm ana} = -8.66$. Therefore, we conclude that the
thermodynamic integration method is rather good in obtaining the
correct evidence of the model. The Laplace approximation
Eq.~(\ref{Laplace}) also fares well for uncorrelated distributions,
$\ln E_{\rm toy5}^{\rm Lap} = -8.67$.

We now consider a likelihood function with a correlated covariance
matrix $C_{ij}$, with the same mean values and dispersions as the
previous case, but with significant correlations. The analytic formula
needed, Eq.~(\ref{evidencenabc}), is no longer exact,\footnote{One could
rotate the parameter basis to remove the correlations, but then the
priors wouldn't be top-hats.} and gives $\ln E_{\rm toy5c}^{\rm ana} =
-7.32$. For comparison thermodynamic integration gives $\ln E_{\rm
toy5c}^{\rm num} = -7.28\pm0.06$, again in perfect agreement within
errors.  In this case the Laplace approximation fails significantly,
$\ln E_{\rm toy5c}^{\rm Lap} = -6.89$, the reason being that the correlations 
chosen bring the posterior into significant contact with the edges of the 
priors.

Let us now return to the uncorrelated case and include a new parameter,
$x_6$, as in Table~I, and evaluate the different evidences that appear
because of this new parameter, in order to see the sensitivity to
systematic errors in the evaluation of the Bayesian evidence and their
effects on model comparison. The numerical result is $\ln E_{\rm
toy6}^{\rm num} = -10.75\pm0.03$, while the exact analytical expression
gives $\ln E_{\rm toy6}^{\rm ana} = -10.74$, in perfect agreement,
within errors. The Laplace approximation Eq.~(\ref{Laplace}) again fares
well for uncorrelated distributions, $\ln E_{\rm toy6}^{\rm Lap} =
-10.74$. 

When the likelihood function has large correlations, and the priors
are not too large, the naive Laplace approximation,
Eq.~~(\ref{Laplace}), fares less well than the analytical
approximation, Eq.~~(\ref{evidencenabc}).

\subsection{A real model comparison}

In this subsection we will make use of the results obtained in
Ref.~\cite{Beltran2005}, where we evaluated the evidence for 
5- and 6-parameter adiabatic models, and for three 10-parameter mixed adiabatic 
plus 
isocurvature models.
The prior ranges used are given in Table~II. The latter models give a
marginally better fit to the data but require more parameters, which is
exactly the situation where model selection techniques are needed to
draw robust conclusions. In Ref.~\cite{Beltran2005} we used
thermodynamic integration to compute the evidence and showed that the
isocurvature models ware less favoured than the adiabatic ones, but only at
a mild significance level.\footnote{Recently Trotta \cite{Trotta} used a 
different technique to analyze a restricted class of isocurvature model 
featuring just one extra parameter, and found it highly disfavoured. The 
different conclusion is primarily due to the very different prior he chose on 
the isocurvature amplitude, such that almost all the models under the prior are 
domintaed by isocurvature modes and in poor agreement with the data.}

Beginning with the simplest adiabtic model, which uses the Harrison--Zel'dovich 
spectrum, we have used the analytical formulae
above, Eq.~(\ref{evidencenabc}), together with the covariance matrix
provided by the {\tt cosmoMC} programme \cite{cosmomc}, and obtained $\ln E_{\rm
ad}^{\rm ana} = -854.07$, while the thermodynamical integration gave
$\ln E_{\rm ad}^{\rm num} = -854.1\pm0.1$ \cite{Beltran2005}. The agreement is 
excellent; this is because the distribution function for the
adiabatic model is rather well approximated by a Gaussian, and the
priors are rather large, so the formula Eq.~(\ref{evidencenabc}) is very close
to that obtained in the Laplace approximation, $\ln E_{\rm ad}^{\rm Lap}
= -854.08$.

\begin{table}[thb]
\begin{center}
\begin{tabular}{c c c c}
\hline Parameter &\hspace{5mm} Mean&\hspace{1cm}Prior Range\hspace{5mm}
&\hspace{3mm}Model\hspace{3mm}\\ 
\hline
$\omega_{\rm b}$
&\hspace{5mm}0.022&\hspace{5mm}[0.018,\,0.032]&\hspace{5mm}AD-HZ,AD-$n_{{\rm 
s}}$,ISO \\
$\omega_{\rm dm}$ 
&\hspace{5mm}0.12&\hspace{5mm}[0.04,\,0.16]&\hspace{5mm}AD-HZ,AD-$n_{{\rm 
s}}$,ISO \\
$\theta$ &\hspace{5mm}1.04&\hspace{5mm}[0.98,\,1.10]
&\hspace{5mm}AD-HZ,AD-$n_{{\rm s}}$,ISO \\
$\tau$ &\hspace{5mm}0.17&\hspace{5mm}[0,\,0.5]
&\hspace{5mm}AD-HZ,AD-$n_{{\rm s}}$,ISO \\
$\ln[10^{10}{\Rrad}]$ &\hspace{5mm}3.1&\hspace{5mm}[2.6,\,4.2]
&\hspace{5mm}AD-HZ,AD-$n_{{\rm s}}$,ISO \\
$n_{\rm s}$ &\hspace{5mm}1.0&\hspace{5mm}[0.8,\,1.2]
&\hspace{5mm}AD-$n_{{\rm s}}$,ISO \\
$n_{\rm iso}$ &\hspace{5mm}1.5&\hspace{5mm}[0,\,3] &\hspace{5mm}ISO \\
$\delta_{\rm cor}$ &\hspace{5mm}1.5&\hspace{5mm}[$-$0.14,\,0.4]
&\hspace{5mm}ISO \\
$\sqrt{\alpha}$ &\hspace{5mm}0&\hspace{5mm}[$-$1,\,1] &\hspace{5mm}ISO \\
$\beta$ &\hspace{5mm}0&\hspace{5mm}[$-$1,\,1] &\hspace{5mm}ISO \\
\hline
\end{tabular}
\end{center}
\caption{\label{params} The parameters used in the models; see 
Ref.~\cite{Beltran2005} for nomenclature and other details. For the
AD-HZ model $n_{\rm s}$ was fixed to $1$ and $n_{\rm iso}$,
$\delta_{{\rm cor}}$, $\alpha$ and $\beta$ were fixed to $0$. In the
AD-$n_{{\rm s}}$ model, $n_{{\rm s}}$ also varies. Every isocurvature
model holds the same priors for the whole set of parameters. }
\end{table}

However the analytic method fares less well for the adiabatic model with
varying $n_{{\rm s}}$, with both the analytic and Laplace methods giving
$\ln E_{{\rm AD-n_{{\rm s}}}} = -853.4$, while the numerical method
gives the smaller value -854.1, a discrepency of nearly unity.

Turning now to the iscurvature cases, we found an extremely good result
for the CDI model, gaining from Eq.~(\ref{evidencenabc}) the value $\ln
E_{\rm cdi}^{\rm ana} = -855.08$, while the thermodynamical integration
gives $\ln E_{\rm cdi}^{\rm num} = -855.1\pm0.1$.  This is surprising,
given the relatively large non-gaussianities for at least three
variables: $n_{\rm iso}$, $\beta$ and $\delta_{\rm cor}$, whose priors
are {\em not} centered with respect to the mean. However the NID case
shows much less good agreement, with a discrepency of 0.6. That suggests
that the closeness of the CDI comparison is to some extent a statistical
fluke, with the underlying method less accurate.

A summary of the different models can be found in Table~\ref{models}.

\begin{table}[thb]
\begin{center}
\begin{tabular}{l c c c c}
\hline Model &\hspace{6mm}$\ln {\cal L}^{\rm max}$
&\hspace{9mm}$\ln E^{\rm num}$&\hspace{8mm}$\ln E^{\rm ana}$
&\hspace{8mm}$\ln E^{\rm Lap}$\\
\hline
toy5 &\hspace{5mm}0
&\hspace{5mm}$-8.65\pm0.03$&\hspace{5mm}$-8.66$&\hspace{5mm}$-8.67$\\
toy5c &\hspace{5mm}0
&\hspace{5mm}$-7.28\pm0.06$&\hspace{5mm}$-7.32$&\hspace{5mm}$-6.89$\\
toy6 &\hspace{5mm}0
&\hspace{3.5mm}$-10.75\pm0.03$&\hspace{3.5mm}$-10.74$&\hspace{3.5mm}$-10.74$\\
toy6c &\hspace{5mm}0
&\hspace{5mm}$-9.73\pm0.06$&\hspace{5mm}$-9.71$&\hspace{5mm}$-9.63$\\
\hline
AD &\hspace{5mm}$-840.78$
&\hspace{5mm}$-854.1\pm0.1$&\hspace{5mm}$-854.1$&\hspace{5mm}$-854.1$\\
AD-$n_{{\rm s}}$ &\hspace{5mm}$-838.50$
&\hspace{5mm}$-854.1\pm0.1$&\hspace{5mm}$-853.4$&\hspace{5mm}$-853.4$\\
CDI &\hspace{5mm}$-838.05$
&\hspace{5mm}$-855.1\pm0.2$&\hspace{5mm}$-855.1$&\hspace{5mm}$-854.5$\\
NID &\hspace{5mm}$-836.60$
&\hspace{5mm}$-855.1\pm0.2$&\hspace{5mm}$-854.5$&\hspace{5mm}$-854.5$\\
NIV &\hspace{5mm}$-842.53$
&\hspace{5mm}$-855.1\pm0.3$&\hspace{5mm}$-854.9$&\hspace{5mm}$-854.9$\\
\hline
\end{tabular}
\end{center}
\caption{\label{models} The different models, both toy and real,
with their maximum likelihoods and evidences.}
\end{table}

\subsection{Savage--Dickey method}

Another numerical method for evidence calculation is the Savage--Dickey
method, first described in Ref.~\cite{SD} and recently used in 
Ref.~\cite{Trotta}. This technique allows one to calculate
the evidence ratio of two models from a simple and quick analysis of
the Markov chains used for parameter estimation, provided that the
models are nested; i.e., that one of them is included in the
parameter space of the other. For instance, the AD model is nested
within the AD-$n_{{\rm s}}$ model, and the AD and AD-$n_{{\rm s}}$ models are 
both nested
within the CDI, NID and NIV ones. In the context of Markov chains, the
Savage--Dickey method is essentially a measure of how much time the
sampler spends in the nested model, weighted by the respective volumes
of the two models. When the outer model has extra parameters, this method
relies on approximating the nested model as a model with negligibly
narrow priors in directions of extra parameters. We note,
however, that when many extra parameters are present, this method must
fail for reasons similar to those why grid-based parameter estimation
approaches fail with models with many parameters. The MCMC parameter
estimation simply does not have high enough dynamic range to probe the
two models given the large prior volume ratio.

The AD and AD-$n_{{\rm s}}$ models differ by one parameter. Using the same AD+ns
samples as for the analytic method (i.e., the samples from which we
extracted the covariance matrix), we obtained $\ln (E_{AD}/E_{AD+n_s})
= 0.03$.  The result from the precise thermodynamical
integration, $\ln (E_{\rm AD}/E_{{\rm AD}-n_{{\rm s}}}) = 0 \pm 0.1$ is in 
excellent
agreement. The AD-$n_{{\rm s}}$ and CDI (or NID, NIV) models differ by four
parameters.  With most simple choices of parametrization (including in
particular the isocurvature and cross-correlation tilts), the AD-$n_{{\rm s}}$
is not a point, but a hypersurface within the parameter space of the
isocurvature models (i.e. $\alpha=0$ and other three parameters act as
dummy, unconstrained, parameters which do not affect the evidence). In
these cases, the evidence ratios given by the Savage--Dickey method do
not converge as the priors of the extra parameters are tightened up
around the nested model, although they match thermodynamically-determined values 
to within a unit of $\ln E$.

\section{Discussion and Conclusions}

We have developed an analytical formalism for computing the Bayesian
evidence in the case of an arbitrary likelihood distribution with a
hierarchy of non-Gaussian corrections, and with arbitrary top-hat
priors, centered or uncentered. This analysis can be of great help for
the problem of model comparison in the present context of cosmology,
where observational data is still unable to rule out most extensions
of the standard model based on the $\Lambda$CDM inflationary paradigm.

As an application of the exact and approximate formulae obtained for
the Bayesian evidence of a model with approximately Gaussian
likelihood distributions, we have compared the value predicted
analytically with that computed with a time-consuming algorithm based
on the thermodynamical integration approach. The values obtained
analytically agree surprisingly well with those obtained
numerically. While one can estimate the magnitude of the higher order
corrections for the analytical formulae, it is very difficult to
estimate the systematic effects of the numerical approach. Thus, with
this analytical method we can test for systematics in the
thermodynamical integration approach.  So far, the values obtained
agree, so it seems that the numerical approach is a good tool for
estimating the evidence. However, it takes considerable effort and
machine time to do the correct evaluation, and therefore, we propose
the use of the analytical estimate, whose corrections are well under
control, in the sense that one can compute the next order corrections
and show that they are small.

\

{\bf Note added:} Many years after my work was finished, a book appeared~\cite{book} which 
thoroughly discussed Bayesian Methods in Cosmology.

\end{document}